\documentstyle[eqsecnum,preprint,aps]{revtex}
\tighten
\draft
\begin{document}
\widetext
\preprint{CLNS 96/1414}
\bigskip
\bigskip
\title{Asymmetric Non-Abelian Orbifolds and Model Building}
\medskip
\author{Zurab Kakushadze\footnote{zurab@hepth.cornell.edu \\ 
\hbox{$\;\;$} Address after September 1, 1996: Lyman Laboratory, 
Harvard University,
Cambridge, MA 02138.},
Gary Shiu\footnote{shiu@hepth.cornell.edu}, 
and S.-H. Henry Tye\footnote{tye@hepth.cornell.edu}}
\bigskip
\address{Newman Laboratory of Nuclear Studies,
Cornell University,
Ithaca, NY 14853-5001, USA}
\date{July 16, 1996}
\bigskip
\medskip
\maketitle
\begin{abstract}
{}The rules for the free fermionic string model construction are extended to 
include general non-abelian orbifold constructions that go beyond
the real fermionic approach. This generalization is also applied to the
asymmetric orbifold rules recently introduced. These non-abelian orbifold 
rules are quite easy to use. Examples are given to illustrate their
applications.
\end{abstract}
\pacs{}

%============================================================================
%local macro
\def\hf{-{1\over 2}}
\def\phf{{1\over 2}}
\def\3hf{{3\over 2}}
\def\5hf{{5\over 2}}
\def\oalpha{\overline{\alpha V}}
\def\obeta{\overline{\beta V}}
\def\HL{H^L_{\oalpha}}
\def\HR{H^R_{\oalpha}}
\def\SGF{V_{i} \cdot N_{\oalpha}~-\sum k_{ij} \alpha_{j}-s_i-V_{i} \cdot 
\oalpha}
\def\bZ{\bf Z}
\def\tilN{\tilde{N}}
\newcommand{\ket}[1]{\left| {#1} \right>}

%%%%%%%%%%%%%%%%%%%%%%

%=============================================================================
\widetext
%\newpage
\section{Introduction}
\bigskip

{}This paper presents the generalization of the rules for the free             
fermionic and asymmetric abelian orbifold constructions 
to include non-abelian orbifolds. The orbifold construction\cite{DHVW}, 
both abelian and non-abelian, is an algebraic approach to string model
building. The rules presented here are particularly useful for asymmetric 
orbifolds\cite{NSV}, where the geometric approach is less powerful.

	Consider a typical compactification of an $n$-dimensional 
space $R^n$. An $n$-dimensional orbifold $\Omega$ is constructed by 
identifying 
points of $R^n$ under a space group $S$ of rotations and translations. 
The subgroup of $S$ formed by pure translations
is referred to as the lattice $\Lambda$, since identification
of points of $R^n$ under $\Lambda$ defines the torus $\Gamma^n$. In this 
case, $\Omega$ can be obtained by identifying points of $\Gamma^n$ under 
the point group $P$, which is the discrete subgroup of only rotations in $S$. 
This may be summarized by
\begin{equation}
\Omega = R^{n}/S = \Gamma^{n}/P ~.
\end{equation}
Of course, the lattice $\Lambda$ in a consistent string model is
Lorentzian. Also, it is understood that appropriate twisted sectors must be
included in the orbifold model. More generally, one may construct the same 
model $\Omega$ by starting with another model $\Omega^{'}$ modded 
(or orbifolded) 
by an appropriate group $G$ of translations and rotations.
In general, there are more than one choice of ($\Omega^{'}$, $G$) that 
will reproduce 
the same $\Omega$. For example, we may start with an appropriate 
torus $\overline{\Gamma}$
and identify its points under a specific group $\overline{P}$ of translations 
and rotations. Here $\overline{P}$ is a subgroup of the isometry group of 
the torus $\overline{\Gamma}$.
\begin{equation}
\Omega = \Omega^{'}/G = \overline{\Gamma}/ \overline{P} = \cdots~.
\end{equation}
{\it A priori}, the group $G$ may be either abelian or non-abelian. 
(Here, a typical $\overline{P}$ is non-abelian.) An orbifold that involves 
a (non-)abelian $G$ is referred to as a (non-)abelian orbifold. 
An asymmetric orbifold refers to an orbifold when the group actions 
on the left movers and on the right movers are different.
So it may happen that the same model $\Omega$ can be reached via an abelian 
orbifold or via a non-abelian orbifold, either symmetric or asymmetric, 
all depending on which $\Omega^{'}$ one starts with.

        Let us recall the situation with the free fermionic string model 
construction\cite{KLT}. A particular free fermionic string model is 
dictated by the choice of spin structures and their correlations. 
Since the spin structures are generically left-right 
asymmetric, these models are asymmetric orbifolds in the orbifold language.
However, the present rules for the free fermionic string construction
do not allow general types of non-abelian orbifolds. This is
unnecessarily restrictive, since in this construction, we must always 
start from a unique $\Omega^{'}$. For example, in the 4-dimensional 
heterotic string case, $\Omega^{'}$ is simply
the $SO(44)$ tachyonic string model. 
Although the set of spin structures is obviously
commuting, in the real world-sheet fermionic basis, they do yield models
that correspond to a special type of non-abelian orbifolds (that involve
non-commuting ${\bf Z}_2$ twists) in the usual orbifold language.
However, the present rules do not allow the construction of other types 
of non-abelian orbifolds. A simple example will illustrate this point.
	
	Consider the ${\bf Z}_2$ orbifold of a single compactified boson, 
which can have arbitrary radius. However, if we use the present free 
fermionic string construction, we must start with the boson at radius 1. 
The rules allow us to construct either 
	
(1) a ${\bf Z}_2$ twisted boson at radius 1 by giving 
the two real world-sheet fermions different spin structures (one 
periodic and one anti-periodic), or 

(2) a boson with radius at any rational value, by giving the single 
complex fermion appropriate spin structures.

But the present free fermionic string rules do not permit the 
construction of a twisted boson at a radius away from 1, since the 
two sets of spin structures mentioned above are not compatible 
with each other. In the orbifold language, the above construction 
involves $G=\overline{P}$ which is non-abelian.

	Fortunately, the generalization of the rules needed is 
not difficult to obtain.
In this paper, we shall extend the free fermionic construction rules 
to include orbifold actions of non-abelian groups. Some of the important 
steps in reaching our rules were already developed in the literature. 
Besides the original papers by Dixon, Harvey, Vafa and 
Witten\cite{DHVW}, we refer the readers to the work by Ginsparg\cite{gins},
and that by Li and Lam\cite{LiLam}. We believe the rules presented in 
this paper are relatively easy to use, especially for intricate models.
With these new rules, we expect the free fermionic construction to be 
closer to the usual orbifold constructions. This 
should not be surprising, since the equivalence of 
world-sheet fermions and bosons is well-known. 

	Recently the free fermionic string construction rules have 
been generalized to include world-sheet bosons\cite{Kaku}. These 
relatively simple rules are useful in the construction of asymmetric 
orbifolds, since they allow us to consider twists other than 
${\bf Z}_2$ twists. In this construction, the starting $\Omega^{'}$ is 
quite flexible. Generically, it is chosen to be an $N=4$ supersymmetric 
Narain model\cite{Narain}. The generalization of these rules to include
non-abelian orbifolds is essentially identical to that for the 
free fermionic case. Using these rules, it is quite straightforward to
construct general non-abelian orbifolds, including models with non-abelian
point groups. The readers who are interested only in the bosonic 
formulation may go 
directly to section V, then to section III and VI.

    We shall start with the free fermionic string construction.
In section II, we briefly review its rules.
In section III, we give the generalized rules for the construction of 
consistent non-abelian orbifold models within the framework of the free 
fermionic string construction.
In section IV, we illustrate these rules by giving some explicit
examples. We consider examples in both ten and four dimensions.
In section V, we review the rules for asymmetric orbifolds.
The generalization of these rules to the non-abelian case is identical
to that given for the free fermionic case in section III.
To illustrate the use of the rules for non-abelian asymmetric orbifold
constructions, we give two examples in section VI.
The first example involves a non-abelian point group $P$, namely the
permutation group $S_3$, plus Wilson lines.
The second example is the three-family $SO(10)_3$ model recently 
constructed\cite{three}.  Here, we recast that model as a non-abelian orbifold.
Appendix A gives a discussion of the ${\bf Z}_2$ orbifold of a single 
compactified boson.

%=============================================================================
\widetext
%\newpage
\section{Preliminaries}

\bigskip

{}In this section, we briefly review the rules for constructing
free fermionic string models as presented in Ref.\cite{KLST}. 
To be concrete, let us focus on heterotic strings 
compactified to four space-time dimensions. In the light cone gauge which we 
adopt, the world-sheet degrees of freedom consist of two string 
coordinates and 10~(22) right~(left) moving chiral world-sheet complex 
fermions. 
A string model is succinctly defined by the choice of spin structures, 
{\it i.e.},
a set of basis vectors $\{V_{0},V_{1},\cdots,V_{n}\}$.
Each $V_i$ is a $32$ (rational) component vector with Lorentzian signature
$((-1)^{10},(+1)^{22})$.
Let $\alpha_{i}V_{i}$ and $\beta_{i}V_{i}$ be two linear combinations that
specify the boundary conditions of the 32 complex fermions 
$\psi^{\ell}$, $\ell=1,2,\cdots, 32$ on the torus,
\begin{eqnarray}
&&\psi^{\ell}(\sigma_{1} + 2 \pi,\sigma_{2}) 
= -e^{-2 \pi i \alpha_{i}V_{i}^{\ell}} \psi^{\ell}(\sigma_{1},\sigma_{2})~, 
\nonumber\\
&&\psi^{\ell}(\sigma_{1},\sigma_{2} + 2 \pi) 
= -e^{-2 \pi i \beta_{i}V_{i}^{\ell}} \psi^{\ell}(\sigma_{1},\sigma_{2})~.
\end{eqnarray}
It is convenient to choose $\hf \leq V_{i}^{\ell} < {1\over 2}$, and we
define $\oalpha=\alpha V - \Delta(\alpha)$ where 
$\Delta^{l}(\alpha) \in {\bf Z}$, so that 
$\hf \leq \overline{\alpha V^{\ell}} < {1\over 2}$.
Consistency requires the presence of 
$V_{0}=(\hf (\hf \hf \hf)^{3}
 \vert (\hf)^{22})$ where the first component labels the boundary
condition of the superpartner of the two right-moving transverse space 
coordinates. 

{}We impose three consistency requirements on the one-loop string 
partition function $Z$:

\noindent (1) World-sheet supersymmetry, {\it i.e.}, the triplet constraint 
on each
vector $V_{i}$
\begin{eqnarray}\label{triplet}
s_{i} \equiv V_{i}^{1} &=&V_{i}^{2+3n} + V_{i}^{3+3n} + V_{i}^{4+3n} 
\quad ({\mbox{mod}~ 1})  \quad n = 0,1,2  \nonumber\\
                       &=&0~{\mbox{or}}~ \hf~. 
\end{eqnarray}
\noindent This ensures the proper boundary conditions for the world-sheet 
supercurrent and hence space-time Lorentz invariance in the covariant
gauge.

\noindent (2) One-loop modular invariance, and

\noindent (3) Physically sensible projection, {\it i.e.}, every space-time 
degree 
of freedom contributes to the partition function with the proper weight, $+1$
for a space-time boson (with ${\alpha s}=0$ (${\mbox{mod}~ 1}$)) and 
$-1$ for a space-time fermion (with ${\alpha s}=-1/2$ (${\mbox{mod}~ 1}$)).
  
A consistent string model is defined by the set $\{V_{0},V_{1},\cdots,V_{n}\}$
and the structure constants $\{k_{ij}:i,j=0,1,\cdots,n \}$ which satisfy
\begin{eqnarray}
k_{ij} + k_{ji} =&V_{i} \cdot V_{j} &\pmod{1}~, \\
    m_{j}k_{ij} =&0                 &\pmod{1}~, \\
k_{ii} + k_{i0} + s_{i} - {1\over 2} V_{i} \cdot V_{i} =&0 
                                    &\pmod{1}~, 
\end{eqnarray}
where all the dot products are defined with Lorentzian signature, {\it i.e.},
left
movers minus right movers. Here, $m_j$ is the smallest positive 
integer such that
$m_{i}V_{i}^{\ell} \in {\bf Z}$ for all $\ell$. 

The world-sheet fermionic contribution to the one-loop partition 
function is then given 
by
\begin{equation}\label{zf}
Z = {1\over {\prod_i m_{i}}} \sum_{\alpha,\beta} {\mbox Tr} 
\Bigl( (-1)^{\alpha s}~q^{\HL}~{\overline q}^{\HR} 
~e^{-2 \pi i \beta_i (\SGF)} \Bigr)
\end{equation}
where $\alpha s \in {\bZ}~({\bZ}+ {1\over 2})$ for space-time bosons
(fermions). The partition function is divided into different sectors labeled
by the vectors $\oalpha$, with hamiltonian $H^L_{\oalpha}$ ($H^R_{\oalpha}$)
for the left (right) movers, and a vector of fermion number operators, 
$N_{\oalpha}$, for the $32$ complex fermions. A precise definition of
$N_{\oalpha}$ is given in Ref.\cite{KLST}. The states that are kept in 
$Z$ satisfy
\begin{equation}
V_i \cdot N_{\oalpha} = \sum_j k_{ij} \alpha_j + s_i - V_i \cdot \oalpha
\pmod{1}~.
\end{equation}

{}To establish a dictionary between the present construction and
the orbifold approach, let us rewrite the partition function as
\begin{equation}\label{projection}
Z =\sum_{\alpha} Tr \Bigl( (-1)^{\alpha s}~q^{\HL}
~{\overline q}^{\HR}
~\prod_{i=0}^{n} P^{i}_{\alpha V} 
\Bigr) 
\end{equation}
where the sum is over all the sectors $\oalpha$ and
\begin{equation}
P^{i}_{\alpha V} = {1\over m_{i}} \sum_{\beta_{i}=0}^{m_{i}-1} 
e^{- 2 \pi i \beta_{i}(\SGF)}
\end{equation}
are projection operators. The states that survive in $Z$ are invariant under 
all the projections $P^{i}_{\alpha V}$.
On the other hand, the partition function of an orbifold by the action of a
discrete abelian group $G$ is
\begin{eqnarray}\label{Zhg1}
Z &=& \sum_{g \in G} {\mbox{Tr}} 
\Bigl( (-1)^{F_{g}}~q^{H^L_g}~{\overline q}^{H^R_g} P_{g} \Bigr)
= {1\over \vert G \vert} \sum_{g,h \in G} Z(g,h) \\
{\mbox{where}} \quad
P_{g} &=& {1\over \vert G \vert} \sum_{h \in G} h ~. 
\end{eqnarray}
Therefore, for each model constructed from complex world-sheet
fermions, we can associate an abelian orbifold with orbifold group
$G = {\bZ}_{m_0} \times {\bZ}_{m_1} \times \cdots \times {\bZ}_{m_n}$ 
~(where ${\bZ}_{m_0}$ is always ${\bZ}_{2}$). 
To make the correspondence more explicit, we denote $V_g = \oalpha$ and
$V_h =\obeta$. The {\bf 0} sector ($\oalpha = 0$) is the identity sector
while all the others are the twisted (or shifted) sectors.

{}As is clear from the rules, given the orbifold group G, there are distinct
choices of the structure constants~(or torsions) $\{k_{ij}\}$, 
corresponding to different choices of projections in the various sectors.
If the boundary conditions are always periodic or antiperiodic, 
the complex fermion may be split into
2 real fermions, which may then have different boundary conditions. When we
allow real fermions with different boundary conditions, the set $\{V_i \}$
must satisfy the cubic constraint
\begin{equation}
4 \sum_{l:real} V_{i}^{l}~V_{j}^{l}~V_{k}^{l} = 0 \pmod{1} \quad 
\mbox{for all}~i,j,k ~.
\end{equation}
This implies that we can find a complex fermion basis for any three vectors. 
Typically, there 
is no complex fermion basis for the complete set $\{V_{i}\}$.
In order to stay within the framework of complex world-sheet fermions,
one has to introduce the notion of non-abelian orbifolds. 
As we shall see, the free
fermionic string model construction covers only a special type of non-abelian
orbifolds. In the next section, we will further generalize the above 
formulation to include other types of non-abelian orbifolds.

%===========================================================================
\widetext
\section{Non-abelian Orbifolds}
\bigskip

{}We are now ready to discuss the general case of non-abelian 
orbifolds\cite{DHVW,gins,LiLam}.
Starting with a consistent string model, whose Hilbert space admits a 
discrete symmetry $G$, modding out the theory by the action of $G$ generally
results in a new string model. As discussed before, the 
consistency conditions for abelian $G$
can be expressed in terms of $\{k_{ij}, V_i \}$. In this section, we will
examine the constraints when $G$ is non-abelian. Our approach follows that
of Ref.\cite{LiLam}. The main improvement is the inclusion of a 
careful treatment of the compactified $6$ dimensions here, and 
the generalization to the bosonic formulation later.

{}It turns out that the consistency constraints can be inferred from the
associated abelian orbifolds $Z(G_{a})$ where $G_{a}$ are abelian subgroups 
of $G$. To see this, let us go back to Eq.~(\ref{Zhg1}). For abelian $G$, the
operator interpretation of Eq.~(\ref{Zhg1}) is clear. The Hilbert space
decomposes into a set of twisted (and/or shifted) sectors labeled by $g$, 
and in each sector, there is a projection onto $G$ invariant states. 
In taking the 
trace, we can always find a basis that is simultaneously diagonal in $H^L_g$
and $H^R_g$ and $h$. The projection in the case of non-abelian $G$ requires
a more careful treatment. In each twisted sector $g$, the sum over 
$h \in G$ includes
only elements that commute with $g$. The group action, however, mixes states
in conjugate sectors. This follows because, if
\begin{equation}
g \psi^ \ell g^{-1} = - e^{-2 \pi i \oalpha^{l}} \psi^{l}~,
\end{equation}
then for any $c \in G$,
\begin{equation}
(cgc^{-1})(c \psi^{\ell} c^{-1})(cgc^{-1})^{-1} 
= - e^{-2 \pi i \oalpha^{l}} (c \psi c^{-1})~.
\end{equation}
As a result,
\begin{equation}\label{c0}
Z(g,h) = Z(cgc^{-1},chc^{-1})~.
\end{equation}
The partition function may be expressed in terms of conjugacy classes $C_{i}$:
\begin{eqnarray}\label{Zhg2}
Z(G) &=& {1\over \vert G \vert} \sum_{g \in G} \sum_{gh=hg}
{\mbox{Tr}} \Bigl( (-1)^{F_g}~q^{H^L_g}~{\overline q}^{H^R_g}~h \Bigr) 
\nonumber \\
&=& \sum_i {1\over \vert N_i \vert} \sum_{h \in N_i} 
{\mbox{Tr}} \Bigl( (-1)^{F_{C_i}}~q^{H^L_{C_{i}}}
~{\overline q}^{H^R_{C_{i}}}~h \Bigr)
\end{eqnarray}
where the group $N_i$ is the stabilizer~(or little) group of the conjugacy
class $C_i$ and is defined only up to conjugation. In general, the summation 
in Eq.~(\ref{Zhg2}) decomposes into
distinct modular orbits, {\it i.e.}, distinct subsets each of which is modular
invariant. However, individual modular orbit does not satisfy physically
sensible projection. In the full summation in Eq.~(\ref{Zhg2}), physically 
sensible projection is explicitly satisfied since in each sector labeled by
$C_i$, the summation over $h$ is a properly normalized projection onto states 
invariant under the stabilizer group $N_i$. To see that modular invariance
is satisfied, let us consider $G_a$ (the maximal abelian subgroups of $G$),
$G_{ab} = G_a \cap G_b$ and $G_{abc} = G_a \cap G_b \cap G_c$ etc. It is
easy to see that $Z(G)$ may be rewritten as a weighted sum of the abelian
orbifolds $Z(G_a), Z(G_{ab}), Z(G_{abc}),\cdots$
\begin{equation}\label{psum}
Z(G) = {1\over \vert G \vert} \Bigl\{\sum_{a} {\vert G_{a} \vert}~Z(G_a) 
-\sum_{a<b} {\vert G_{ab} \vert}~Z(G_{ab}) 
+\sum_{a<b<c} {\vert G_{abc} \vert}~Z(G_{abc}) - \cdots \Bigr\}
\end{equation}
where $Z(G_a) = \sum_{g,h \in G_a} Z(g,h)$. 
Since each of the $Z(G_{a \cdots})$ is a modular invariant partition
function, modular invariance of $Z(G)$ is automatic. 
Notice that the coefficients in Eq.(\ref{psum}) always add up to $1$. 
Each partition function is properly normalized; this guarantees that the
the final partition function $Z(G)$ contains exactly one graviton.

{}Nevertheless, the above reasoning does not imply that all non-abelian 
orbifolds obtained this way are consistent.
Notice that each abelian orbifold $Z(G_{a})$ is specified by 
$\{ k_{ij}^{a},V_i^{a} \}$
~(as well as the others $Z(G_{ab}) \cdots$). 
For a given $G_a$ ({\it i.e.} the set $\{ V_a \}$), there are, in general, 
inequivalent $Z(G_a)$ models depending on the
choices of $\{ k_{ij}^{a} \}$. In order to determine Eq.~(\ref{psum})
precisely, we must find the appropriate $\{ k_{ij}^{a} \}$ for each $Z(G_a)$. 
It turns out that the non-abelian group properties of $G$ impose 
stringent constraints on the choices of $k_{ij}^{a}$. A consistent non-abelian
orbifold model exists only if all the $k_{ij}$'s satisfy all the constraints.
These constraints are divided into four types:

\bigskip
\noindent {\em 1) Between different abelian orbifolds $Z(G_a)$ and $Z(G_b)$}
\smallskip

{}Consider the $g$ twisted sector in $Z(G_a)$. Suppose there is $c \in G$
such that $cgc^{-1}$ is outside $G_a$, say $cgc^{-1}$ belongs to $G_b$. 
Since $g$ and $cgc^{-1}$ belong to the same conjugacy class, we have from
(\ref{c0}), $Z_{a}(g,h) = Z_{b}(cgc^{-1},chc^{-1})$, {\it i.e.}, $Z(G_a)$ and
$Z(G_b)$ are isomorphic. With an appropriate choice of the bases 
$(V_i^a=V_i^b)$, we have the constraints relating $Z(G_a)$ and $Z(G_b)$,
\begin{equation}\label{c1}
k_{ij}^a=k_{ij}^b ~.
\end{equation}
We shall impose this constraint in the outset.

\bigskip
\noindent {\em 2) Within the same abelian orbifold $Z(G_a)$}
\smallskip

{}Next, suppose that the isomorphism is inner, {\it i.e.}, both 
$g, cgc^{-1} \in G_a$.
Let $g_i \in G_a$ be the generators of $G_a$, {\it i.e.},
\begin{eqnarray}
V_{g_i} = V_i \quad {\mbox{and}} \quad 
V_{cgc^{-1}} = {\overline{\lambda_{il} V_{l}}} ~.\nonumber
\end{eqnarray}
Let us consider specific terms in $Z(G_a)$. From Eq.~(\ref{c0}),
$Z_{a}(g,h) = Z_{a}(cgc^{-1},chc^{-1})$, where
\begin{eqnarray*}
V_g = \oalpha ~, \quad \quad V_h= \obeta ~, \quad \quad
V_{cgc^{-1}} = {\overline{\alpha \lambda V}} ~, \quad \quad
V_{chc^{-1}} = {\overline{\beta \lambda V}} ~.
\end{eqnarray*}
Since these two sectors are isomorphic, if we confine ourselves to the
$N_{\oalpha} = N_{\overline{\alpha \lambda V}} = 0$ states in the trace, we
have
\begin{eqnarray*}
e^{2 \pi i \beta_i(\sum k_{ij} \alpha_{j} + s_i - V_i \cdot \oalpha)}
= e^{2 \pi \beta_i \lambda^i_l(\sum k_{ln} \lambda^j_n \alpha_j + s_i 
-V_l \cdot \overline{\alpha \lambda V})} ~. 
\end{eqnarray*}
Choosing $V_g=V_j$, $V_h=V_i$, we obtain the constraint within 
$\{ k_{ij}, V_i \}_a$,
\begin{equation}\label{c2}
k_{ij} =\lambda_{il} k_{ln} \lambda_{nj} - \Delta_i \cdot \lambda_{jn} V_n
\pmod{1}
\end{equation}
where $\Delta_i = \lambda_{il} V_l - {\overline{\lambda_{il} V_l}}$.
Notice that $s_i=\lambda_{il} s_l \!\pmod{1}$ since conjugation does not mix 
space-time bosons and fermions.

\bigskip
\noindent {\em 3) Between $Z(G_a)$ and $Z(G_{ab})$}
\smallskip

{}Then, we need to find the relations between $\{ k_{ij}, V_{i} \}_a$ and
$\{ k_{ln}, V_l \}_{ab}$ where $G_{ab}$ is a subgroup of $G_a$. Clearly,
we can write, for each generator $g \in G_{ab}$,
\begin{eqnarray*}
V_l^{ab} = \overline{\gamma_{li} V_i^a} 
\end{eqnarray*}
and
\begin{eqnarray}
k_{ln}^{ab} + k_{nl}^{ab} = V_l^{ab} \cdot V_n^{ab} 
&=&\overline{\gamma_{li} V_i^a} \cdot \overline{\gamma_{nj} V_j^a} \nonumber\\
&=&(\gamma_l V^a - \Delta_l)(\gamma_n V^a - \Delta_n) \nonumber\\
&=& \gamma_l^i \gamma_n^j (k_{ij}^a + k_{ji}^a) 
- \Delta_{l} \cdot \gamma_n V^a - \Delta_{n} \cdot \gamma_l V^a  ~.\nonumber
\end{eqnarray}
Consistency with (\ref{c2}) gives
\begin{equation}\label{c3}
k_{ln}^{ab} = \gamma_l^i k_{ij}^a \gamma_n^j - \Delta_l \cdot \gamma_n V^a~.
\end{equation}

\bigskip
\noindent {\em 4) Between three different abelian orbifolds $Z(G_a)$, 
$Z(G_b)$ and $Z(G_c)$}
\smallskip

{}Consider an element $g$ that is in $G_a$, $G_b$ and $G_c$. Suppose 
$h_{1} \in G_a$, $h_{2} \in G_b$. To obtain non-trivial constraints, 
$h_{1}h_{2} \in \!\!\!\!\!/~G_a,G_b$, {\it i.e.}, $[h_{1}h_{2},h_{1}] \neq 0$,
$[h_{1}h_{2}, h_{2}] \neq 0$. Let $h_{1}h_{2} \in G_c$ and 
$V_g =\overline{\alpha^{a}V^a} = \overline{\alpha^{b}V^b} 
=\overline{\alpha^{c} V^c}$. Although the same $g$ twisted 
sector is in $Z(G_a)$, $Z(G_b)$ and $Z(G_c)$, it is expressed in different 
bases.
Fortunately, the operators that relate these different bases commute with
$q^{H^L_g} \overline{q}^{H^R_g}$. If we restrict ourselves to the 
$N_{\overline{\alpha^{a}V^a}} = N_{\overline{\alpha^{b}V^b}} 
=N_{\overline{\alpha^{c} V^c}} = 0$ 
states, the projection operators $h_{1}$, $h_{2}$ and $h_{1}h_{2}$
are purely phases, and their group relation yields,
\begin{eqnarray*}
e^{2 \pi i \bigl[\beta^{a}_i(V^a_i \cdot \overline{\alpha^{a}V^a} - 
\sum k_{ij}^a \alpha_{j}^{b} -s_i^a)
+ \beta^{b}_i(V^b_i \cdot \overline{\alpha^{b}V^b} -
\sum k_{ij}^b \alpha_{j}^{b} -s_i^b) \bigr]}
=e^{2 \pi i \beta^{c}_i(V^c_i \cdot \overline{\alpha^{c}V^c} -
\sum k_{ij}^c \alpha_{j} -s_i^c)} ~.
\end{eqnarray*}
Since $\beta^{a}s^a + \beta^{b}s^b = \beta^{c} s^c \pmod{1}$, we have the
relation between $\{k_{ij},V_i \}_a$, $\{ k_{ij}, V_i \}_b$ and
$\{ k_{ij}, V_i \}_c$,
\begin{equation}\label{h1}
\beta^{a}_i(V^a_i \cdot \overline{\alpha^{a} V^a} - k_{ij}^a \alpha_j^{a})+
\beta^{b}_i(V^b_i \cdot \overline{\alpha^{b}V^b} - k_{ij}^b \alpha_j^{b})
=\beta^{c}_i(V^c_i \cdot \overline{\alpha^{c} V^c} - k_{ij}^c \alpha_j) 
\pmod{1}
\end{equation}
where $V_g= \overline{\alpha^{a}V^a}= \overline{\alpha^{b}V^b}
= \overline{\alpha^{c} V^c}$ and
$V_{h_1}= \overline{\beta^{a}V^{a}}$,
$V_{h_2}= \overline{\beta^{b}V^{b}}$,
$V_{h_{1}h_{2}}= \overline{\beta^{c}V^{c}}$. In the examples given
in the next section, we shall see that Eq.~(\ref{h1}) is a stringent 
constraint for consistent non-abelian orbifolds, especially when the 
non-abelian orbifold is also asymmetric.

\medskip

{}Let us summarize the steps required to construct consistent non-abelian 
orbifolds: \\
(1) Rewrite the partition function $Z$ as a sum of abelian orbifold partition
functions (Eq.~(\ref{psum})).\\
(2) Check that the free fermionic construction 
rules are satisfied in each abelian orbifold $\{k_{ij},V_i \}_A$. \\
(3) Impose extra constraints (\ref{c1}),(\ref{c2}),(\ref{c3}) and 
(\ref{h1}) between various $\{ k_{ij}, V_i \}_A$.
With an appropriate choice of the basis vectors, Eq.~(\ref{c1}) is 
satisfied automatically.

%============================================================================
\widetext
\section{Examples}
\bigskip

{}In this section, we demonstrate how to apply our rules by 
constructing some explicit examples. Before we focus on 
four-dimensional models, let us begin our discussion with 
ten-dimensional examples that exhibit some of the basic features 
of our construction.

\subsection{Ten-dimensional Models}
{}There are only nine known consistent heterotic string models in 
ten dimensions\cite{DH}. Among them, eight have rank 16 gauge group and 
can be realized in terms of complex world-sheet fermions. The ninth has gauge
group a single $E_8$ realized at level $2$, and was first constructed using
{\it real\/} fermion basis. In the bosonic formulation, this higher
level model can be obtained by modding out the $E_{8} \times E_{8}$ 
model by the combined action of a $2 \pi$ spatial rotation and a ${\bZ}_2$ 
permutation of the two $E_8$'s\cite{GinsVafa}. We now illustrate how the 
same model can be constructed by a $D_4$ orbifold using the rules just given.
We are using this well-known example to clarify our notation.

{}The group $D_4$ is generated by two non-commuting elements $r$ and $\theta$ 
with the defining relations,
\begin{equation}
r^2 = \theta ^{4} = 1,  \quad \quad \quad r \theta^{3}= \theta r ~.
\end{equation}
It has 8 elements and is divided into 5 conjugacy classes: 
$C_{1}=\{1\}$, $C_{\theta^{2}}=\{\theta^{2}\}$, $C_{r}=\{r, r \theta^{2}\}$,
$C_{r \theta}=\{r \theta, r \theta^{3}\}$, and 
$C_{\theta}=\{\theta, \theta^{3}\}$.
The stabilizer groups are given by 
$N_{1}=N_{\theta ^{2}}=D_4$,
$N_{r}=\{1, \theta^{2}, r, r \theta^{2}\}$,
$N_{r \theta}=\{1, \theta^{2}, r \theta, r \theta^{3}\}$, and
$N_{\theta}=\{1, \theta, \theta^{2}, \theta^{3}, \theta^{4}\}$.
It is straightforward to express the full partition function as a
sum of abelian orbifolds,
\begin{equation}\label{D4}
Z = \phf~Z_{\theta} + \phf~Z_{r} + \phf~Z_{r \theta} - \phf~Z_{\theta^{2}}
\end{equation}
where $Z_{\theta}$ is a ${\bZ}_{4}$ ( $\cong$ $N_{\theta}$) orbifold, 
$Z_{r}$ and
$Z_{r \theta}$ are ${\bZ}_{2} \times {\bZ}_{2}$ ( $\cong$ $N_{r}$ and
$N_{r \theta}$ respectively) orbifolds, $Z_{\theta^{2}}$ is a ${\bZ}_{2}$ 
($\cong$ $\{1, \theta^{2}\}$) orbifold.

{}We have not specified the representations of the internal fermions under 
the action of $D_4$. There are 5 irreducible 
representations of $D_4$: 4 one-dimensional 
representations defined by $\theta=\pm 1$, $r=\pm 1$ and only one 
two-dimensional representation:  
\begin{equation}
\theta = \left( \begin{array}{cc}
                i & 0  \\
                0 & -i
                \end{array} 
         \right) ~,
\quad \quad 
     r = \left( \begin{array}{cc}
                0 & 1  \\
                1 & 0
                \end{array}
         \right) 
\end{equation}
where we have chosen the basis such that $\theta$ is diagonal.

{}Let us start with the simplest model, {\it i.e.}, the model generated by a 
single
basis vector $V_0=((\hf)^{4} \mid (\hf)^{16})$. This model is 
non-supersymmetric with 32 tachyons and gauge group $SO(32)$. 
We may define the $D_4$ action by embedding eight copies of the 
two-dimensional representation on the left movers while leaving the right 
movers untouched. In other words, the left movers transform under $\theta$
and $r$ as follows:
\begin{eqnarray}
\theta: \quad \quad
\psi^{2 \ell -1} &\rightarrow& \mbox{\space \space} i~ \psi^{2 \ell -1} 
\nonumber \\
\psi^{2 \ell}    &\rightarrow& -i~ \psi^{2 \ell}   \nonumber \\
r: \quad \quad
\psi^{2 \ell -1} &\rightarrow& \psi^{2 \ell}     \\
\psi^{2 \ell}    &\rightarrow& \psi^{2 \ell-1} \nonumber
\end{eqnarray}
for $\ell=1,\cdots,8$. The generating vectors when expressed in their own 
diagonal bases are:
\begin{eqnarray}
V_{\theta}&=& \Bigl( 0^{4} \mid ({1\over 4} -\!\!{1\over 4})^{8} \Bigr)~,
\nonumber \\
V_{r}&=& \Bigl( 0^{4} \mid (\hf~0)^{8} \Bigr) ~,\nonumber \\
V_{r \theta}&=& \Bigl( 0^{4} \mid (\hf~0)^{8} \Bigr) ~,  \\
V_{\theta^{2}}&=& \Bigl( 0^{4} \mid (\hf)^{16} \Bigr) ~.\nonumber
\end{eqnarray}

{}We now examine the abelian models in some detail. To begin, let us consider
the $Z_r$ model generated by the abelian set $\{ V_0, V_{\theta^{2}}, V_r \}$.
The matrix of dot products $V_i \cdot V_j$ and the structure constants 
$k_{ij}$ are given by:
\begin{equation}
 V_i \cdot V_j = \left( \begin{array}{ccc}
                         3 & 4 & 2 \\
                         4 & 4 & 2 \\
                         2 & 2 & 2
                        \end{array}
                 \right) ~,
\quad \quad
         k_{ij}= \left( \begin{array}{ccc}
                         k_{00}         &k_{\theta^{2}0} & k_{r0} \\
                         k_{\theta^{2}0}&k_{\theta^{2}0} & k_{r \theta^{2}} \\
                         k_{r0}         &k_{r \theta^{2}} & k_{r0}
                        \end{array}
                 \right) ~,
\end{equation}
where $k_{ij} = 0,\phf \pmod{1}$. There are six potentially massless sectors:
${\bf 0}$, $V_{0}+V_{\theta^{2}}$,
$V_r$, $V_{r}+V_{\theta^{2}}$, $V_{0}+V_{r}$ and $V_{0}+V_{r}+V_{\theta^{2}}$.
The spectrum generating formula in the ${\bf 0}$ sector (with vacuum energy
$[\hf,-1]$) reads,
\begin{eqnarray}
\phf~\sum_{i=1}^{4} N_{i} + \phf~\sum_{i=1}^{16} \tilN_{i} &=& \phf \pmod{1}
~, \nonumber \\
\phf~\sum_{i=1}^{16} \tilN_{i} &=& 0 \pmod{1} ~, \\
\phf~\sum_{i=1}^{8} \tilN_{2i-1} &=& 0 \pmod{1} ~. \nonumber
\end{eqnarray}
The tachyons are projected out and the lowest energy states include the 
graviton and gauge bosons in the adjoint of $SO(16) \times SO(16)$. 
On the other hand, the $V_{0}+V_{\theta^{2}}$ sector 
(with vacuum energy $[0,-1]$) provides the accompanying superpartners if
and only if $k_{r0}=k_{r \theta^{2}}$. This follows from the constraints,
\begin{eqnarray}
\phf~\sum_{i=1}^{4} N_{i} + \phf~\sum_{i=1}^{16} \tilN_{i} =&
k_{00}+k_{\theta^{2}0}+\phf &\pmod{1} ~, \nonumber \\
\phf~\sum_{i=1}^{16} \tilN_{i} =& 0 & \pmod{1} ~, \\
\phf~\sum_{i=1}^{8} \tilN_{2i-1} =& k_{r0}+k_{r \theta^{2}} &\pmod{1} ~.
\nonumber 
\end{eqnarray}
The gravitino ($\tilN_{i}=0$ for all i) survives the projection if 
$k_{r0}=k_{r \theta^{2}}$. In addition, we have space-time fermions in the
adjoint representation of $SO(16) \times SO(16)$. If $k_{r0} \neq 
k_{r \theta^{2}}$,
the gravitino is projected out and the space-time fermions form 
the $({\bf 16},{\bf 16})$ representation of $SO(16) \times SO(16)$.

{}There may be additional gauge bosons coming from the sectors $V_r$ and 
$V_{r}+V_{\theta^{2}}$. In the $V_r$ sector (with vacuum energy
$[\hf,0]$), the spectrum generating formula gives,
\begin{eqnarray}
\phf~\sum_{i=1}^{4} N_{i} + \phf~\sum_{i=1}^{16} \tilN_{i} 
=& k_{r0} + \phf &\pmod{1} ~, \nonumber\\
\phf~\sum_{i=1}^{16} \tilN_{i}=& k_{r \theta^{2}} &\pmod{1} ~, \\
\phf~\sum_{i=1}^{8} \tilN_{2i-1} =& k_{r0} &\pmod{1} ~. \nonumber
\end{eqnarray}    
Provided that $k_{r0}=k_{r \theta^{2}}$, there are extra
gauge bosons in the ${\bf 128}$ spinor representation of the first $SO(16)$. 
The $V_{r}+V_{\theta^{2}}$ sector differs from the $V_r$ sector by an 
interchange of left moving NS and R fermions. Therefore, it 
provides gauge bosons in the spinor representation of the second $SO(16)$. 
Together with the gauge 
bosons from the ${\bf 0}$ sector, they furnish the adjoint representation of 
$E_8 \times E_8$.

{}Finally, the sectors $V_{0}+V_{r}$ and $V_{0}+V_{r}+V_{\theta^{2}}$ 
contribute space-time fermions in the $({\bf 128},{\bf 1})$ 
and $({\bf 1},{\bf 128})$ representation of $SO(16) \times SO(16)$ 
for all choices of $k_{ij}$.

{}To summarize, modding out the non-supersymmetric $SO(32)$ model 
by ${\bZ}_2 \times {\bZ}_2$ may result in two
completely different models: a supersymmetric $E_8 \times E_8$ model 
if $k_{r0} =k_{r \theta^{2}}$ and a non-supersymmetric 
$SO(16) \times SO(16)$ model otherwise.

{}The above analysis also applies to the $Z_{r \theta}$ model as it has
the same generating vectors (although in different bases). 
The $Z_{\theta^{2}}$ model is generated by $\{ V_{0}, V_{\theta^{2}} \}$
and so there are only two constraints in each sector. The ${\bf 0}$ 
sector provides the graviton and gauge bosons in the adjoint of $SO(32)$
while their superpartners reside in the $V_{0}+V_{1}$ sector. 
 
{}What remains is $Z_{\theta}$ generated by $\{ V_0,V_{\theta} \}$. The matrix
of dot product $V_i \cdot V_j$ and the structure constants $k_{ij}$ are:
\begin{equation}
 V_i \cdot V_j = \left( \begin{array}{cc}
                          3 & 0 \\
                          0 & 1
                        \end{array}
                 \right) ~,
\quad \quad
        k_{ij} = \left( \begin{array}{cc}
                          k_{00}      & k_{\theta 0} \\
                          k_{\theta 0}& k_{\theta 0} + \phf
                        \end{array}
                 \right) ~,
\end{equation}
where $k_{ij}=0,\phf \pmod{1}$. There are 6 potentially massless sectors: 
${\bf 0}$, $V_{0}+2V_{\theta}$, $V_{\theta}$, $3V_{\theta}$, 
$V_{0}+V_{\theta}$ and $V_{0}+3V_{\theta}$. The ${\bf 0}$ sector 
contributes 256 gauge bosons in the adjoint representation of $U(16)$. 
The sectors $V_{0}+V_{\theta}$ and $V_{0}+3V_{\theta}$
together provide 240 space-time fermions.
If $k_{\theta 0}=0$, the $V_{0}+2V_{\theta}$ sector contributes the 
superpartners
of the ${\bf 0}$ sector. In addition, there are extra gauge bosons coming
from the sectors $V_{\theta}$ and $3V_{\theta}$ (both with vacuum energy
$[\hf,\hf]$) that complete the adjoint
representation of $SO(32)$. As a result, we simply recover the 
$Z_{\theta^{2}}$ model.
However, if $k_{\theta 0}=\phf$, the gravitino is projected out and 
there are 240 space-time fermions from the $V_{0}+2V_{\theta}$ sector. 
The massless states in the sectors $V_{\theta}$ and $3V_{\theta}$ are 
projected out but the 2 tachyons survive the projection. The 
model has gauge group $U(16)$.

{}What is the resultant $D_4$ orbifold? {\it A priori}, there are 
6 possibilities corresponding to different choices of $k_{ij}$ in 
$Z_r$, $Z_{r \theta}$ and $Z_{\theta}$ (after taking into account the 
fact that $Z_r \cong Z_{r \theta}$). However, not all of
them are consistent. The non-abelian rules further restrict the number of 
possibilities to three. First consider the constraints
between conjugate sectors $V_r$ and 
$V_{r \theta^{2}}=\overline{V_{r}+V_{r \theta^{2}}}$.
>From definition, $\lambda_{rr}=\lambda_{r \theta^{2}}=1$ and 
$\Delta_{r}=V_{r}+V_{\theta^{2}}-V_{r \theta^{2}}=(0^{4} \mid (-1)^{8}~0^{8})$
>From Eq.~(\ref{c2}), 
\begin{eqnarray}
k_{rr} &=& k_{rr} + k_{r \theta^{2}} + k_{\theta^{2} r} + 
          k_{\theta^{2} \theta^{2}} - \Delta_{r} \cdot (V_{r}+V_{\theta^{2}})
~, \nonumber \\
k_{rj} &=& k_{rj} + k_{\theta^{2}j} - \Delta_{r} \cdot V_j ~, \\
k_{jr} &=& k_{jr} + k_{j \theta^{2}} ~, \nonumber
\end{eqnarray}
where $j=0, \theta^{2}$. It follows that the above equations give rise to 
a single constraint,
\begin{equation}
k_{\theta^{2}0}=0 ~.
\end{equation} 

{}The other conjugate sectors are $\{ V_{r \theta}, V_{r \theta^{3}} \}$
and $\{ V_{\theta}, V_{\theta^{3}} \}$. The former gives the same constraint
as above. In the latter case, notice that 
$V_{\theta^{3}}=\overline{3V_{\theta}}$. Therefore, 
$\lambda_{\theta \theta}=3$ and $\Delta_{\theta}=(0^{4} \mid 1^{8}(-1)^{8})$.
The constraints become,
\begin{eqnarray}
k_{\theta \theta} &=& 9 k_{\theta \theta} -\Delta_{\theta} \cdot 3V_{\theta}
                   = 9 k_{\theta \theta} ~, \nonumber\\
k_{\theta 0}      &=& 3 k_{\theta 0} -\Delta_{\theta} \cdot V_{0} 
                   = 3 k_{\theta 0}   ~,   \\
k_{0 \theta}      &=& 3 k_{0 \theta} ~, \nonumber
\end{eqnarray}
which are automatically satisfied.

{}We now turn to the constraints between three different abelian orbifolds. 
Let us take $V_{h_1}=V_{r}$, $V_{h_2}=V_{\theta}$ and
$V_{h_{1}h_{2}}=V_{r \theta}$. The common sector $V_g$ can be either
$V_0$ or $V_{\theta^{2}}$. From Eq.~(\ref{h1}),
\begin{eqnarray}
V_{r} \cdot V_0 - k_{r0} + V_{\theta} \cdot V_{0} - k_{\theta 0} 
&=& V_{r \theta} \cdot V_{0} - k_{r \theta, 0} ~, \nonumber \\
V_{r} \cdot V_{\theta^{2}} - k_{r \theta^{2}} + V_{\theta} \cdot 
\overline{2V_{\theta}} - 2 k_{\theta \theta}
&=& V_{r \theta} \cdot V_{\theta^{2}} - k_{r \theta, \theta^{2}} ~,
\end{eqnarray}
which give
\begin{eqnarray}\label{constraints}
k_{r0}+k_{\theta 0} &=& k_{r \theta, 0} ~, \nonumber \\
k_{r \theta^{2}} &=& k_{r \theta, \theta^{2}} ~.
\end{eqnarray}

{}As a consequence, the parameters $k_{ij}$ in the abelian models are not
independently chosen. If $k_{\theta 0}=0$ ({\it i.e.} $Z_{\theta}$ is the 
$SO(32)$ model), Eq.~(\ref{constraints}) implies
that $Z_r$ and $Z_{r \theta}$ both have gauge group $E_8 \times E_8$ or 
$SO(16) \times SO(16)$. It is obvious from Eq.~(\ref{D4}) that the resultant 
$D_4$ orbifold is either the $E_8 \times E_8$ or the 
$SO(16) \times SO(16)$ model. 
On the other hand, if $k_{\theta 0}=\phf$ ({\it i.e.}
$Z_{\theta}$ is the $U(16)$ model), $Z_r$ and $Z_{r \theta}$ have gauge
group $E_8 \times E_8$ and $SO(16) \times SO(16)$ respectively. From 
Eq.~(\ref{D4}),
\begin{equation}
Z= \phf~Z_{E_8 \times E_8} + \phf~Z_{SO(16) \times SO(16)} +
   \phf~Z_{U(16)} - \phf~Z_{SO(32)} ~.
\end{equation}
By simply counting the number of gauge bosons (248), space-time fermions (496)
and tachyon (1) in the partition function $Z$, one can easily identify 
the resultant $D_4$ orbifold
as the only rank 8 model in ten dimensions with gauge group $E_8$ 
realized at level 2. This can be confirmed by a careful construction of
the physical states.

{}We have seen how the rules provide non-trivial constraints between the
abelian orbifolds. There are cases in which the constraints cannot be
satisfied and as a result, no consistent model can be obtained. 
Let us take the quaternion group $Q$ as an example\cite{LiLam}. 
It is generated by two order 4 elements p and q ($p \neq q$) with the 
defining relations,
\begin{equation}
p^4=q^4=1, \quad \quad p^2=q^2, \quad \quad qp=p^3q ~.
\end{equation}
There are 5 conjugacy classes: $C_{1}=\{1\}$, $C_{p^{2}}=\{ p^2 \}$,
$C_{p}=\{ p, p^3 \}$, $C_{q}=\{ q, q^3 \}$ and $C_{pq}=\{pq, (pq)^3\}$.
The stabilizer groups are $N_{1}=Q$, $N_{p^2}=Q$, $N_{p}=\{1, p, p^2, p^3\}$,
$N_{q}=\{1, q, q^2, q^3 \}$ and $N_{pq}=\{1, pq, (pq)^2, (pq)^3 \}$. 
Therefore, the full partition function is 
\begin{equation}
Z= \phf~Z_{p} + \phf~Z_{q} + \phf~Z_{pq} - \phf~Z_{p^2} ~.
\end{equation}
The one-dimensional representations are the same as that of $D_4$ while the
two-dimensional representation is
\begin{equation}
p= \left( \begin{array}{rr}
              i & 0 \\
              0 & -i
          \end{array}
   \right) ~,
\quad \quad
q= \left( \begin{array}{cc}
              0 & i \\
              i & 0
          \end{array}
    \right) 
\end{equation}
where we have chosen the basis such that $p$ is diagonal. Both $q$ and $pq$
take the same diagonal form as that of $p$ in their diagonal bases. 
Therefore, $Z_p$, $Z_q$ and $Z_{pq}$ are isomorphic and in what follows, we
will only consider $Z_p$.

{}Again, we start with the model generated by $V_0$. We may embed four copies 
of the two-dimensional representation on the first eight left movers, leaving
all the other internal fermions untouched. The generating vector $V_p$ is
thus
\begin{equation}
V_p = \Bigl( 0^4 \mid ({1\over 4}-\!\!{1\over 4})^{4}~0^{8} \Bigr) ~.
\end{equation}
The matrix of dot products $V_i \cdot V_j$ and the structure constants
$k_{ij}$ of $Z_p$ are given by
\begin{equation}
 V_i \cdot V_j = \left( \begin{array}{cc}
                          3 & 0 \\
                          0 & \phf
                         \end{array}
                 \right) ~,
\quad \quad
        k_{ij} = \left( \begin{array}{cc}
                          k_{00}    &  k_{p0}           \\
                          k_{p0}    &  k_{p0}+{1\over 4}
                        \end{array}
                 \right) ~,
\end{equation}
where $k_{00}, k_{p0}= 0,\phf \pmod{1}$.
The group multiplication (Eq.~(\ref{h1})) is not satisfied. To see this, take 
$g=p^2$, $h_1=p$, $h_2=q$ and $h_{1}h_{2}=pq$. On one hand,
\begin{equation}
V_{p} \cdot 2V_{p} - 2 k_{pp} + V_{q} \cdot 2V_{q} - 2 k_{qq} = 0 ~,
\end{equation}
but on the other hand,
\begin{equation}
V_{pq} \cdot 2V_{pq} - 2 k_{pq,pq} = \phf ~.
\end{equation}
However, it is straightforward to see that if we embed eight copies of the 
two-dimensional representation on the left movers, {\it i.e.},
\begin{equation}
V_p = \Bigl( 0^4 \mid ({1\over4}-\!\!{1\over 4})^8 \Bigr) ~,
\end{equation}
there are two consistent models via this $Q$ orbifold: the supersymmetric 
$SO(32)$ model and the $U(16)$ model. 

{}In a similar fashion, one can show that the constraint Eq.~(\ref{c2}) is
not satisfied when the orbifold group is the extra-special $p$-group (for any 
odd $p\geq5$)\cite{LiLam,freed}.

\subsection{Four-dimensional Models} 
{}We have taken our exercise on ten-dimensional examples far enough, let us
proceed to the more interesting four-dimensional models. A new ingredient
in four dimensions is the world-sheet supersymmetry which is encoded in
the triplet constraint (Eq.~(\ref{triplet})).
 
{}Let us start with a free fermionic string model defined by the set of 
basis vectors:
\begin{eqnarray}
V_0 &=& \Bigl( \hf~(\hf \hf \hf)^3 \mid (\hf)^{22} \Bigr) ~, \nonumber \\
V_1 &=& \Bigl( \hf~(\hf~0~0)^3 \mid 0^{22} \Bigr) ~, \nonumber \\
V_2 &=& \Bigl( \hf~(\hf~0~0) (0 \hf~0)^{2} \mid (\hf)^{10}~0^{10}~(\hf)^{2}
\Bigr) ~, \\
V_3 &=& \Bigl( 0^{7}~(0 \hf \hf) \mid 0^{10} (\hf)^2~0^{5}~(\hf)^{4}~0 \Bigr)
~. \nonumber
\end{eqnarray}

{}The matrix $V_{i} \cdot V_{j}$ and the structure constants
$k_{ij}$ are given by,
\begin{equation}
 V_i \cdot V_j = \left( \begin{array}{rrrr}
                        3 & -1   & 2      & 1       \\
                       -1 & -1   & \hf   & 0        \\
                        2 & \hf & 2      & 0        \\
                        1 & 0    & 0      & 1       \\
                        \end{array}
                \right) ~,
\quad \quad
 k_{ij}        = \left( \begin{array}{cccc}
                        0&0     &0          &0     \\
                        0&0     &k_{21}+\phf&k_{31}\\
                        0&k_{21}&\phf       &k_{32}\\
                        0&k_{31}&k_{32}     &\phf  \\
                        \end{array}
                \right) ~.
\end{equation}

{}In general, the particle content depends on the choices of $k_{ij}$. In
this model, however, $\{k_{0j}:j=0, \cdots, 3\}$ do not affect the massless
spectrum and we may simply set them to zero.
$V_1$ introduces gravitinos and hence space-time supersymmetry. $V_2$ serves
the dual purpose of cutting the supersymmetry from $N=4$ to $N=2$, and the 
gauge group from $SO(44)$ to $[SO(20)]^{2} \times SO(4)$. 
$V_3$ breaks the gauge group further to 
$SO(20) \times SO(4) \times SO(10) \times SO(6) \times [U(1)]^2$. If
$k_{31} = {1\over 2}$, $V_3$ removes all the gravitinos 
, otherwise $k_{31}=0$ and the model remains $N=2$ 
supersymmetric.

{}Taking the N=2 supersymmetric model ({\it i.e.}, choosing $k_{31}=0$) as our 
starting point, we can
reduce the supersymmetry to $N=1$ as well as obtain a higher level gauge
group via non-abelian orbifold. Again, take $D_4$ as an example. 
We define the $D_4$ action by choosing the following generating vectors,
\begin{eqnarray}
V_{\theta}&=&\Bigl(\hf~(0 \hf~0)(0 \hf~0)(\hf~0~0) \mid 
({1\over 4}-\!\!{1\over 4})^{8}(\hf)^{2}~0^{2}(\hf)^{2} \Bigr) ~, \nonumber \\
V_{r}     &=&\Bigl(\hf~(0 \hf~0)(\hf~0~0)(0~0 \hf) \mid (\hf~0)^{8}~0^{6}
\Bigr) ~, \nonumber \\
V_{r \theta}&=&\Bigl(~0^{4}~(\hf \hf~0)(\hf~0 \hf) \mid (\hf~0)^{8}(\hf)^{2}
~0^{2} (\hf)^{2} \Bigr) ~, \\
V_{\theta^{2}}&=&\Bigl(~0^{10} \mid (\hf)^{16}~0^{6} \Bigr) ~. \nonumber
\end{eqnarray}
In other words, we assign two-dimensional representations only on the first
16 left movers and one-dimensional representations on the others. 

{}Before we work out the complete spectrum, let us gain some insights
from the individual abelian models. $V_{\theta^{2}}$ does not break 
supersymmetry and so $Z_{\theta^{2}}$ remains $N=2$. On the other hand, 
both $V_r$ and $V_{\theta}$ break the supersymmetry to $N=1$. 
Using simple counting 
argument, physically sensible projection for the gravitinos can only be 
satisfied if $Z_{r \theta}$ has $N=0$ or $2$. As we shall see, $V_{r \theta}$
either removes all the gravitinos or leaves them untouched, depending on 
whether $k_{ r \theta 1}=k_{21}$.
 
{}Chiral fermions can appear only in $N=1$ models. {\it A priori}, 
both $Z_{\theta}$
and $Z_r$ may contain chiral fermions. However, in order for $Z_{\theta}$
to be part of the $D_4$ orbifold, $Z_{\theta}$ is necessarily non-chiral.
To see this, let us assume that $Z_{\theta}$ is chiral. A particle from the
$\theta$ sector together with an antiparticle from the $\theta^{3}$ sector 
form a single chiral multiplet. On the other hand, $\theta$ and $\theta^{3}$
belong to the same conjugacy class and hence half of the states from these two
sectors are projected out. Physically sensible projection is not
satisfied because the chiral multiplet cannot be divided into half.

{}Let us examine the abelian models in more detail. First consider
$Z_{\theta^{2}}$ generated by the vectors 
$\{ V_{0}, \cdots, V_{3}, V_{\theta^{2}} \}$. 
The matrix of dot products $V_{i} \cdot V_{j}$ 
and the structure constants $k_{ij}$ are given by:
\begin{equation}
 V_i \cdot V_j = \left( \begin{array}{ccccc}
                        \cdot&       &       &      & 4        \\
                             & \cdot &       &      & 0        \\
                             &       & \cdot &      & \5hf    \\
                             &       &       & \cdot& \phf      \\
                           4 &  0    & \5hf  & \phf & 4
                        \end{array}
                \right) ~,
\quad \quad
 k_{ij}        = \left( \begin{array}{ccccc}
                        \cdot&       &      &     & k_{\theta^{2} 0}     \\
                             & \cdot &      &     & k_{\theta^{2} 1}     \\
                             &       & \cdot&     & k_{\theta^{2} 2}+\phf \\
                             &       &      &\cdot& k_{\theta^{2} 3}+\phf \\
        k_{\theta^{2} 0}&k_{\theta^{2} 1}&k_{\theta^{2} 2}&k_{\theta^{2} 3}
                        &k_{\theta^{2} 0}
                        \end{array}
                \right) ~.
\end{equation}
The gauge bosons only come from the ${\bf 0}$ sector. With the addition of
$V_{\theta^{2}}$, the gauge group is broken to 
$SO(20) \times SO(4) \times SO(8) \times U(1) \times SO(6) \times [U(1)]^{2}$.
The 2 gravitinos in the $V_1$ sector are not projected out
and the model remains $N=2$ supersymmetric.

{}Next, $Z_r$ is generated by 
$\{ V_{0}, \cdots, V_{3}, V_{\theta^{2}}, V_{r} \}$ and therefore
\begin{equation}
 V_i \cdot V_j = \left( \begin{array}{rrrrrr}
                        \cdot&       &       &      &       & 1    \\
                             & \cdot &       &      &       & \hf  \\
                             &       & \cdot &      &       & 1    \\
                             &       &       & \cdot&       & 0    \\
                             &       &       &      & \cdot & 2    \\ 
                           1 &  \hf  & 1     & 0    & 2     & 1
                         \end{array}
                \right) ~,
\quad \quad
 k_{ij}= \left( \begin{array}{cccccc}
               \cdot&      &      &       &                & k_{r0}\\
                    &\cdot &      &       &                & k_{r1+\phf}\\
                    &      &\cdot &       &                & k_{r2} \\
                    &      &      &\cdot  &                &k_{r3} \\
                    &      &      &       &\cdot           &k_{r \theta^{2}} \\
              k_{r0}&k_{r1}&k_{r2}&k_{r3} &k_{r \theta^{2}}&k_{r0}
                \end{array}
        \right) ~.
\end{equation}
The gauge group is broken further by $V_{r}$ to 
$[SO(10)]^{2} \times [U(1)]^{2} \times [SO(4)]^{4} \times U(1) \times SO(6)
\times [U(1)]^{2}$. The supersymmetry is reduced to $N=1$ due to the extra
constraint:
\begin{equation}
V_{r} \cdot N_{V_1} = \phf~( N_1 + N_5 ) = k_{r1} + \phf - V_{r} \cdot V_1
=k_{r1} ~.
\end{equation}
The $V_r$ sector has vacuum energy $[0,0]$. The spectrum generating formula
reads,
\begin{eqnarray}
\phf~N_1 + \phf~(\tilN_1 + \tilN_3 + \cdots + 
\tilN_9 )  =& k_{2r} - \phf &\pmod{1} ~, \nonumber \\
\phf~(\tilN_1 + \tilN_3 + \cdots \tilN_{15}) =& k_{\theta^{2}r} 
&\pmod{1} ~, \nonumber \\
\phf~(N_1 + N_5) =& k_{1r} &\pmod{1} ~, \\
\phf~(N_{10} + \tilN_{11}) =& k_{3r} &\pmod{1} ~, \nonumber \\
\phf~(N_1 + N_3 + N_5 + N_{10}) =& k_{\theta^{2}r} + k_{0r} - \phf &\pmod{1}
~. \nonumber
\end{eqnarray}
It contributes chiral fermions in the representation 
$({\bf 16},{\bf 1},{\bf 2},{\bf 1}) + 
({\bf 16},{\bf 1},\overline{\bf 2},{\bf 1})$ 
or $(\overline{\bf 16},{\bf 1},{\bf 2},{\bf 1}) + 
(\overline{\bf 16},{\bf 1},\overline{\bf 2},{\bf 1})$ of 
$[SO(10)]^{2} \times [SO(4)]^{2}$ (the first two $SO(4)$) depending on the
space-time helicity. The $V_{r}+V_{\theta^{2}}$ sector differs from the
$V_r$ sector by an interchange of the first 16 left moving NS and R fermions
and so contributes chiral fermions in the representation 
$({\bf 1},{\bf 16},{\bf 1},{\bf 2}) +
({\bf 1},{\bf 16},{\bf 1},\overline{\bf 2})$
or $({\bf 1},\overline{\bf 16},{\bf 1},{\bf 2}) +
({\bf 1},\overline{\bf 16},{\bf 1},\overline{\bf 2})$. Altogether, there 
are eight families of chiral fermions in the 
${\bf 16}$ (or $\overline{\bf 16}$) of $SO(10)$.

{}The analysis of $Z_{r \theta}$ is similar and we shall be brief. With the 
generating vectors $\{ V_{0}, \cdots, V_{3}, V_{\theta^{2}}, V_{r\theta} \}$,
\begin{equation}
 V_i \cdot V_j = \left( \begin{array}{rrrrrr}
                        \cdot&       &       &      &       & 2    \\
                             & \cdot &       &      &       & \hf  \\
                             &       & \cdot &      &       & \3hf    \\
                             &       &       & \cdot&       & \phf   \\
                             &       &       &      & \cdot & 2    \\ 
                           2 & \hf   & \3hf  & \phf & 2     & 2
                         \end{array}
                \right) ~,
\quad \quad
 k_{ij}= \left( \begin{array}{cccccc}
          \cdot&      &      &       &                &k_{r\theta,0}\\
               &\cdot &      &       &                &k_{r\theta,1}+\phf\\
               &      &\cdot &       &                &k_{r\theta,2}+\phf\\
               &      &      &\cdot  &                &k_{r\theta,3}+\phf\\
               &      &      &       &\cdot           &k_{r\theta,\theta^{2}}\\
  k_{r\theta,0}&k_{r\theta,1}&k_{r\theta,2}&k_{r\theta,3}
               &k_{r\theta,\theta^{2}} &k_{r\theta,0}
                \end{array}
        \right) ~.
\end{equation}
Again, gauge bosons only come from the ${\bf 0}$ sector and with the
addition of $V_{r \theta}$, the gauge group is broken to
$[SO(10)]^{2} \times [U(1)]^{2} \times [SO(4)]^{2} \times [U(1)]^{2}
\times SO(4) \times [U(1)]^{2}$. The 2 gravitinos in the $V_1$ sector 
are either all projected out (if $k_{r \theta,1} \neq k_{21}$) 
or remain untouched (if $k_{r\theta,1}=k_{21}$).

{}We finally come to $Z_{\theta}$ which is generated by 
$\{V_0, \cdots, V_{3}, V_{\theta} \}$. Therefore,
\begin{equation}
 V_i \cdot V_j = \left( \begin{array}{rrrrr}
                        \cdot&       &       &      & 0        \\
                             & \cdot &       &      & \hf      \\
                             &       & \cdot &      & 0        \\
                             &       &       & \cdot& \phf     \\
                           0 &  \hf  & 0     & \phf & 1
                        \end{array}
                \right) ~,
\quad \quad
 k_{ij}        = \left( \begin{array}{ccccc}
                        \cdot&       &      &     & k_{\theta 0}       \\
                             & \cdot &      &     & k_{\theta 1}+\phf  \\
                             &       & \cdot&     & k_{\theta 2}       \\
                             &       &      &\cdot& k_{\theta 3}+\phf  \\
                  k_{\theta0}&k_{\theta1}&k_{\theta2}&k_{\theta3} &k_{\theta0}
                        \end{array}
                \right) ~.
\end{equation}
The gauge group is broken to $U(10) \times U(2) \times U(4) \times 
[U(1)]^{2} \times SO(4) \times [U(1)]^{2}$. In contrast to $V_r$, $V_{\theta}$
breaks the supersymmetry to $N=1$ but does not introduce chiral fermions.

{}What can we say about the resultant $D_4$ orbifold simply
by counting? If we choose $k_{r \theta,1}=k_{21}$, there are 2 gravitinos in 
$Z_{r \theta}$ and the resultant model has $N=1$ supersymmetry. In addition,
there are 68 gauge bosons and four families of chiral fermions in the 
${\bf 16}$ of $SO(10)$. 

{}A thorough examination of the spectrum (such as identification of the 
gauge group) requires a more careful treatment. There are 27 potentially 
massless sectors, 13 of which are projected out by the
physically sensible projection. The model is supersymmetric and so it 
suffices to list only half of the remaining sectors (the others can be 
obtained by adding $V_1$). The untwisted sectors include:
${\bf 0}$, $V_{2}+V_{\theta^{2}}$ and $V_{2}+V_{3}+V_{\theta^{2}}$ 
while the twisted sectors include: $V_{r}$, $V_{r}+V_{\theta^{2}}$,
$V_{\theta}$ and $3V_{\theta}$.

{}The 68 gauge bosons all come from the ${\bf 0}$ sector. In the basis such
that $\theta$ is diagonal, the spectrum generating formula gives,
\begin{eqnarray}
\phf~(N_1 + \cdots + N_{10})+ \phf~(\tilN_{1} + \cdots + \tilN_{22})
 &=& \phf   \pmod{1} ~,\nonumber \\
\phf~(N_1 + N_2 + N_5 + N_8) &=& \phf \nonumber \pmod{1} ~,\nonumber \\
\phf~(N_1 + N_2 + N_6 + N_9) + \phf~(\tilN_1 + \cdots + \tilN_{10} 
  + \tilN_{21} + \tilN_{22}) &=& \phf \pmod{1} ~,\\
\phf~(N_9 + N_{10}) + \phf~(\tilN_{11} + \tilN_{12} + \tilN_{18} + \cdots
  + \tilN_{21}) &=& 0 \pmod{1} ~,\nonumber \\
\phf~(N_1 + N_3 + N_6 + N_8) + {1\over 4}~(\tilN_1 - \tilN_2 + \cdots
  + \tilN_{15} - \tilN_{16}) &&\nonumber \\ 
+ \phf~(\tilN_{17} + \tilN_{18} + \tilN_{21}
  + \tilN_{22}) &=& \phf \pmod{1} ~. \nonumber
\end{eqnarray}
The states that survive the projection are invariant under $\theta$.
As mentioned before, the gauge bosons form the adjoint representation of 
$U(10) \times U(2) \times U(4) \times [U(1)]^{2} \times SO(4) \times U(1)^{2}$.
Take the $U(10)$ bosons for example. They are obtained by exciting 
$\tilN_{i}= (\delta_{ij}-\delta_{ik})$ for 
$j,k= 1,3,\cdots,9$ or $2,4,\cdots,10$ or 
$\tilN_{i}=\pm (\delta_{ij} + \delta_{ik})$ for 
$j=1,3,\cdots,9$ and $k=2,4,\cdots,10$.

{}These states, however, are not invariant under the full non-abelian group
$D_4$. Under the action of $r$, $\tilN_{2i-1}$ and $\tilN_{2i}$ for 
$i=1,\cdots 8$ are interchanged. Nevertheless, we can obtain group invariant 
states by forming linear combinations of $\theta$ invariant states. 
In other words, 
let $\ket{\theta}$ be an $\theta$ invariant state, then the combination
$\ket{\theta} + r \ket{\theta}$ is invariant under $D_4$. It follows that
the gauge bosons invariant under $D_4$ form the adjoint representation of
$SO(10)_{2} \times U(1) \times SO(4)_{2} \times [U(1)]^{2} \times SO(4)
\times [U(1)]^{2}$. The rank of the gauge group is reduced to 14.
It can easily be verified that the $SO(10)$ and the $SO(4)$ 
are realized at level 2 by examining the OPE of their Kac-Moody currents. 
We may regard the $SO(10)_{2}$ as the observable sector and the rest as
the hidden sector. In what follows, we will only consider states that are
not singlets under $SO(10)$.

{}Besides gauge bosons, the ${\bf 0}$ sector also contributes scalars. 
For example take $N_{i}=\pm \delta_{i2}$ and $\tilN_{i}$ the same as that
of the gauge bosons, we have two families of scalars in the ${\bf 45}$ 
of $SO(10)$. In addition, there are scalars in the $({\bf 10},{\bf 4})$ of
$SO(10) \times SO(4)$.

{}The analysis of the twisted sectors is similar except that group 
invariance requires mixing of states from conjugate sectors. To illustrate
this, consider a physical state in the $V_r$ sector. It is invariant under
$r$ and $\theta^{2}$ and we denote it as $\ket{r,\theta^{2}}$. The
group invariant combination is thus,
\begin{equation}
\ket{r,\theta^{2}} + \theta \ket{r,\theta^{2}} ~. 
\end{equation}
Here $\theta \ket{r, \theta^{2}}$ is in the conjugate sector 
$V_{r}+V_{\theta^{2}}$ because $\theta r \theta^{3} = r \theta^{2}$.
As a result, the conjugate sectors together contribute four families of 
chiral fermions in the ${\bf 16}$ of $SO(10)$.

{}The other sectors only contribute $SO(10)$ singlets. To summarize, 
the resultant model has observable gauge group $SO(10)$ realized at 
level 2, $N=1$ supersymmetry, four chiral families of $SO(10)$ and massless
scalars in the adjoint of $SO(10)$.

{}So far we have limited ourselves to only one-dimensional representations 
on the right movers. With the $[SU(2)]^{6}$ form of the supercurrent,
\begin{equation}
T_F = i \psi^{u} \partial X_{u} + i \sum_{\ell=1}^{6} \psi_1^{\ell}
\psi_2^{\ell} \psi_3^{\ell} ~,
\end{equation}
a general automorphism that respects world-sheet supersymmetry is a 
product of an inner automorphism for the individual $SU(2)$ algebras and
an outer automorphism that permutes the different $SU(2)$. This allows
a truly non-abelian action on the right movers since in general, two such
automorphisms do not commute.

{}Consider two non-commuting automorphisms $\theta$ and $r$ of $[SU(2)]^{2}$
defined by
\begin{eqnarray}
\theta: \quad \quad 
       \psi_j^{1} &\rightarrow& \psi_j^{2}             \nonumber\\
       \psi_j^{2} &\rightarrow& \psi_j^{1} \sigma_j    \nonumber\\
r: \quad \quad
       \psi_j^{1} &\rightarrow& \psi_j^{2}                      \\
       \psi_j^{2} &\rightarrow& \psi_j^{1}             \nonumber
\end{eqnarray}
where $\sigma_1=\sigma_2=-1$, $\sigma_3=1$. It is obvious that 
$\theta^4=r^2=1$ and $r \theta^{3}=\theta r$. Therefore, the group 
generated by $\theta$ and $r$ is $D_4$. 

{}The eigenstates and the corresponding eigenvalues of $\theta$ are:
\begin{eqnarray}
\chi_r^1 =& \psi_3^{1} + \psi_3^{2} ~,
\quad \quad \quad &\theta=1~;  \nonumber\\
\chi_r^2 =& -\psi_3^{1} +\psi_3^{2} ~,
\quad \quad \quad &\theta=-1~; \nonumber\\
\chi^3   =& i \psi_1^{1} - \psi_1^{2} ~,
\quad \quad \quad &\theta=-i  ~;         \\
\chi^4   =& -i \psi_2^{1} - \psi_2^{2} ~,
\quad \quad \quad &\theta=i   ~; \nonumber 
\end{eqnarray}
where $\chi$ with the subscript $r$ are real fermions. 
Similarly, for $r$:
\begin{eqnarray}
\chi_r^1  =& \psi_3^{1} +\psi_3^{2}  ~,
\quad \quad \quad &r=1 ~;\nonumber \\
\chi_r^{2}=& -\psi_3^{1}+\psi_3^{2}  ~,
\quad \quad \quad &r=-1 ~; \nonumber \\
\chi^3    =& \psi_1^{1} + \psi_1^{2} ~,
\quad \quad \quad &r=1  ~;          \\
\chi^4    =& -\psi_2^{1}+ \psi_2^{2} ~,
\quad \quad \quad &r=-1 ~. \nonumber
\end{eqnarray}

{}To illustrate how to construct non-abelian orbifolds with non-trivial 
$SU(2)$ outer automorphisms, let us consider a symmetric $D_4$ orbifold 
of the model generated by $\{ V_0, V_1 \}$. The $D_4$ action is 
specified by the generating vectors:  
\begin{eqnarray}
V_{\theta}&=&\Bigl(\hf~\{0_r, (\hf)_r,-{1\over 4},{1\over 4} \}^{2} 
                       ~(0 \hf~0) \mid 
                   \hf~\{0_r, (\hf)_r,-{1\over 4},{1\over 4} \}^{2} 
                      ~(0 \hf~0) ~0^{12} 
             \Bigr) ~,      \nonumber \\
V_r       &=&\Bigl(\hf~\{0_r,(\hf)_r,0,\hf \}^{2}~(0 \hf~0) 
                         \mid 
                   \hf~\{0_r,(\hf)_r,0,\hf \}^{2}~(0 \hf~0) ~0^{12}
             \Bigr) ~.      
\end{eqnarray}
In other words, two copies of $[SU(2)]^{2}$ transform under $\theta$ and $r$
defined above. The remaining fermions are simply one-dimensional 
representations of $D_4$. 
 
{}In order that the abelian models are well defined, the original set of
generating vectors must commute with the group action. It is easy to
see that $V_0$ and $V_1$ commute with both $V_{\theta}$ and $V_r$. 
The non-abelian rules 
impose constraints on the structure constants: 
\begin{eqnarray}
k_{\theta^{2}j}&=&0 ~, \nonumber \\
k_{rj}+k_{\theta j}&=&k_{r \theta,j} ~, \\
k_{r \theta^{2}}&=&k_{r \theta, \theta^{2}}  ~, \nonumber
\end{eqnarray}
where $j=0,1$. It follows immediately from $k_{\theta^{2}1}=0$ that
$Z_{\theta^{2}}$ remains $N=4$ supersymmetric. The gauge group is broken to
$SO(8) \times SO(36)$. Both $Z_r$ and $Z_{\theta}$ have $N=2$
supersymmetry with gauge groups 
$SO(4) \times SO(4) \times SO(6) \times SO(30)$ and 
$SO(4) \times SO(4) \times SO(36)$ respectively. 
$Z_{r \theta}$ has gauge group $U(4) \times SO(6) \times SO(30)$. 
Depending on $k_{r\theta,1}$, the $N=4$ supersymmetry in $Z_{r\theta}$
is unbroken ($k_{r\theta,1}=0$) or completely broken ($k_{r\theta,1}=\phf$).
As usual, rank reduction and higher level gauge group come with the 
non-abelian orbifold. The resultant $D_4$ orbifold has rank 20 gauge group
$SO(4)_2 \times SO(6) \times SO(30)$ and $N=2$ (or $0$) supersymmetry.
 
{}Similar analysis can be carried out for non-abelian groups 
generated by other automorphisms of $[SU(2)]^{6}$. We may also start with
a model generated by more basis vectors, provided that they commute with 
the automorphisms. A complete classification of the $[SU(2)]^{6}$ 
automorphisms as well as the sets of mutually commuting basis vectors 
have been studied in Ref.\cite{Dreiner}.

\section{Bosonic Formulation}

{}The above rules can be easily generalized to the general orbifold
case. This generalization is easiest if we start with the recent 
rules for asymmetric orbifold\cite{Kaku}, since they are brought to a 
form very similar to that for the free fermionic string construction. 
Because the generalization in this formulation is completely analogous to 
that for the free fermionic construction, we shall be brief.

{}Let us first review the abelian orbifold case.
Again, let us consider heterotic strings compactified to four
space-time dimensions. In the light-cone gauge, which we adopt, we have
the following world-sheet degrees of freedom:
One complex boson $\phi^0$ (corresponding to two transverse
space-time coordinates); three right-moving complex bosons
$\phi^\ell_R$, $\ell=1,2,3$ (corresponding to six internal coordinates);
four right-moving complex fermions $\psi^r$, $r=0,1,2,3$
($\psi^0$ is the world-sheet superpartner of the right-moving component of
$\phi^0$, whereas $\psi^\ell$ are the world-sheet superpartners of
$\phi^\ell_R$, $\ell=1,2,3$);
eleven left-moving complex bosons $\phi^\ell_L$,
$\ell=4,5,...,14$ (corresponding to twenty-two internal coordinates).
Before orbifolding, the
corresponding string model has $N=4$ space-time supersymmetry and the
internal momenta span an even self-dual Lorentzian lattice $\Gamma^{6,22}$.

{}It is convenient to organize the string states into sectors labeled by the
monodromies of the string degrees of freedom. 
These monodromies can be combined into a single vector.
For each complex boson we will specify
either $T^\ell_i$~(twist), or $U^\ell_i$~(shift).
To keep track of whether a given entry
corresponds to a twist or a shift, it is convenient to
introduce {\em auxiliary\/} vectors
\begin{equation}
 W_i=(0 (0~W^1_i) (0~W^2_i) (0~W^2_i)
 \vert\vert
 W^4_i  \cdots W^{14}_i ) ~.
\end{equation}
The double vertical line separates the right- and left-movers.
The entries $W^\ell_i$ are defined as follows: $W^\ell_i={1\over 2}$ if
$T^\ell_i \not=0$; $W^\ell_i
=0$, otherwise. For example,
\begin{eqnarray}
 V_i &=&(V^0_i (V^1_i ~T^1_i)(V^2_i ~T^2_i)(V^3_i ~T^3_i)\vert\vert
 U^4_i \cdots U^{13}_i ~T^{14}_i) ~,\\
 W_i &=&(0 (0~{1\over 2})^3 \vert\vert 0^{10}~\phf) ~,
\end{eqnarray}
where $T^1_i$, $T^2_i$, $T^3_i$ and $T^{14}_i$ correspond to twists,
$U^4_i$,...,$U^{13}_i$ correspond to shifts, and $V^r_i$, $r=0,1,2,3$,
specify the fermionic spin structures.
Without loss of generality we can restrict the values of $V^r_i$ and
$T^\ell_i$ as follows: $-{1\over 2} \leq
V^r_i < {1\over 2}$; $0\leq T^\ell_i <1$
(A complex boson (fermion) with boundary condition
$T^\ell_i~(V^r_i)=0$ or $1\over 2$ can be split into two real bosons
(fermions)). The shifts $U^\ell_i$ can be
combined into a real $(6,22)$ dimensional Lorentzian
vector ${\vec U}_i$ defined up to
the identification ${\vec U}_i \sim {\vec U}_i+{\vec P}$,
where ${\vec P}$ is an arbitrary vector of $\Gamma^{6,22}$.

{}The notation we have introduced proves convenient in describing the
sectors of a given string model based on the orbifold group $G$. 
For $G$ to be a finite abelian group, the element $g(V_i)$ 
must have a finite order $m_i \in {\bf N}$,
{\em i.e.} $g^{m_i} (V_i)=1$. This implies that $V^r_i$ and $T^\ell_i$ must be
rational numbers, and the shift vector ${\vec U}_i$ must be a rational
multiple of a vector in $\Gamma^{6,22}$; that is, $m_i V^r_i , m_i T^\ell_i
\in {\bf Z}$, and $m_i {\vec U}_i \in \Gamma^{6,22}$. To describe all the
elements of the group $G$, it is convenient to introduce the set of
generating vectors $\{ V_i \}$ such that
${\overline {\alpha V}}={\bf 0}$ if and only if
$\alpha_i \equiv 0$. Here ${\bf 0}$ is the null vector:
${\bf 0}=(0 (0~0)^3 \vert\vert 0^{11})$.
Also, $\alpha V \equiv \sum_i \alpha_i V_i$
(the summation is defined as $(V_i +V_j )^\ell=V^\ell_i +
V^\ell_j$), $\alpha_i$ being integers that
take values from $0$ to $m_i -1$. The overbar notation is defined as follows:
${\overline {\alpha V}} \equiv \alpha V -\Delta(\alpha)$, and the components
of ${\overline {\alpha V}}$ satisfy
$-{1\over 2}\leq {\overline {\alpha V}}^{\, r} <{1\over 2}$,
$0\leq {\overline {\alpha T}}^{, \ell}<1$;
here $\Delta^r (\alpha),\Delta^\ell (\alpha)
\in {\bf Z}$. So the elements of the group $G$ are in one-to-one
correspondence with the vectors ${\overline {\alpha V}}$ and will be denoted
by $g({\overline {\alpha V}}$). It is precisely the abelian nature of $G$ that
allows this correspondence (by simply taking all possible linear
combinations of the generating vectors $V_i$).
So the sectors of the model are labeled by the vectors
${\overline {\alpha V}}$. To maintain world-sheet supersymmetry, each 
vector must satisfy the following {\em supercurrent} constraint
\begin{equation}\label{supercurrent}
 V^\ell_i + T^\ell_i =V^0_i \equiv s_i ~(\mbox{mod}~1)~,~~~\ell=1,2,3~.
\end{equation}
Here $s_i$ is the monodromy of the supercurrent
${\overline S}({\overline z}e^{-2\pi i} )=\exp(2\pi is_i)
{\overline S}({\overline z})$, which must satisfy $s_i \in {1\over 2}{\bf Z}$.
Then the sectors with ${\overline {\alpha V}}^{\, 0} =0$
give rise to space-time bosons, while
the sectors with ${\overline {\alpha V}}^{\, 0}
=-{1\over 2}$ give rise to space-time fermions. This is the bosonic 
equivalence of the triplet constraint (\ref{triplet}).

{}The rules for model building is quite similar to that for 
the free fermionic string construction. To obtain relatively simple rules 
for building asymmetric orbifold models,
we will confine our attention to the orbifolds
with twists whose orders are co-prime numbers.
(The  order of a twist generated by a vector ${\overline {\alpha V}}$
is defined as
the smallest positive integer $t({\overline {\alpha V}})$, such that
$\forall\ell~t({\overline {\alpha V}}) {\overline {\alpha T}}^{\,\ell}
\in {\bf Z}$; note that $t({\overline {\alpha V}})$ is a divisor of
$m({\overline {\alpha V}}))$.
In this case, a given model can be
viewed as being generated by a single twist $V^*$ of order $t^*=\prod_i
t_i$, such that ${\overline {\alpha V}}=
{\overline {\alpha^* V^*}}$,
where $\alpha^* /t^* =\sum_i \alpha_i /t_i~({\mbox{mod}~1})$,
and $\alpha^*$ takes values $0,1,...,t^* -1$.
We may work in the $V^*$ basis or the $\{V_i \}$ basis.

{}In a given sector ${\overline {\alpha V}}$, the right- and left-moving
Hamiltonians are given by the corresponding sums of the Hamiltonians for
individual string degrees of freedom. The Hilbert space in the
${\overline {\alpha V}}$ sector is given by
the momentum states $\vert {\vec P}_
{\overline {\alpha V}} +\alpha {\vec U}, {\bf n} \rangle$, and also
the states obtained from these states by acting with the fermion and boson
creation operators (oscillator excitations). ${\bf n}$ is a collective
notation for $n_i=0,1,...,t_i -1$.
In the untwisted sectors, that
is, sectors ${\overline {\alpha V}}$ with $t({\overline {\alpha V}})=1$,
we have ${\vec P}_
{\overline {\alpha V}} \in \Gamma^{6,22}$. In the twisted sectors
${\overline {\alpha V}}$ with $t({\overline {\alpha V}})
\not=1$, we have ${\vec P}_
{\overline {\alpha V}} \in
{\tilde I}({\overline {\alpha V}})$, where ${\tilde I}({\overline {\alpha V}})$
is the lattice dual to the lattice $I({\overline {\alpha V}})$, which
in turn is the sublattice of $\Gamma^{6,22}$ invariant under the action of
the twist part of the group element $g({\overline {\alpha V}})$.
The ground states $\vert {\vec 0}, {\bf n} \rangle$ in the
${\overline {\alpha V}}$ sector
appear with certain degeneracies $\xi({\overline {\alpha V}}, {\bf n})$,
which are non-negative integers.
In the untwisted sectors $\xi({\overline {\alpha V}}, {\bf n}) =1$
if all $n_i=0$, and zero, otherwise.
In the twisted sectors, the
situation is more involved. Here, the 
states with quantum numbers ${\bf n}$
appear with the multiplicity $\xi ({\overline {\alpha V}}, {\bf n})$),
with fixed point phase $f_i ({\overline {\alpha V}}, {\bf n})$ given by
\begin{equation}\label{fixedpointphases}
 f_i ({\overline {\alpha V}}, {\bf n}) =n_i /t_i ~,
\end{equation}
The requirements of modular
invariance and physically sensible projection unambiguously
fix the degeneracies $\xi({\overline {\alpha V}},{\bf n})$.
Now, the sum of $\xi({\overline {\alpha V}}, {\bf n})$
over all ${\bf n}$ is the number
of fixed points $\xi({\overline {\alpha V}})$
in the ${\overline {\alpha V}}$ sector \cite{NSV}:
\begin{equation}
 \sum_{\bf n} \xi({\overline {\alpha V}},{\bf n}) =
\xi({\overline {\alpha V}}) =
 (M({\overline {\alpha V}}))^{-1/2} \prod_\ell[2\sin
 (\pi {\overline {\alpha T}}^{\, \ell})] ~, 
\end{equation}
where the product over $\ell$ does not include the terms with
${\overline {\alpha T}}^{\, \ell} =0$, and $M({\overline {\alpha V}})$ is 
the determinant of the metric of $I({\overline {\alpha V}})$.
We also have the following constraints,
\begin{equation}
\xi({\overline {\alpha V}}) 
\geq
\sum_{\bf n}
 \xi({\overline {\alpha V}},{\bf n}) \exp(-2\pi i \beta
 f({\overline {\alpha V}},{\bf n}) )
\end{equation}
where the sum must be a non-zero integer.
These constraints essentially fix all the 
$\xi({\overline {\alpha V}},{\bf n})$.

{}Since the momentum
in the twisted sectors belongs to a shifted
${\tilde I}({\overline {\alpha V}})$ lattice, the level matching
condition requires that this lattice must have a prime
$N({\overline {\alpha V}})$, where $N({\overline {\alpha V}})$
is the smallest positive integer such that
for all vectors ${\vec P}\in {\tilde I}({\overline {\alpha V}})$,
$N({\overline {\alpha V}}) {\vec P}^2\in 2{\bf Z}$;
moreover, for the corresponding characters to have the correct modular
transformation properties ({\em i.e.}, so that they are permuted under both
$S$- and $T$-transformations),
either $N ({\overline {\alpha V}}) =1$ (in which
case $I({\overline {\alpha V}})$ is an even self-dual lattice), or
$N({\overline {\alpha V}})=
t({\overline {\alpha V}})$ (in which case $I({\overline {\alpha V}})$ is
even but not self-dual).

{}Imposing modular invariance and physical sensible projections, 
the following constraints on a consistent model emerge:	
\begin{eqnarray}
 \label{k1z}
 k_{ij} +k_{ji} =& V_i \cdot V_j &\pmod{1}~,~~~i\not= j ~, \nonumber \\
 k_{ii} +k_{i0} +s_i -{1\over 2} V_i \cdot V_i =&0 &\pmod{1} ~, \nonumber \\
 (k_{ij}-W_i \cdot V_j)m_j =&0 &\pmod{1} ~, \\
 m_i V_i \cdot (W({\overline {\alpha V}}) -\sum_j \alpha_j W_ j)
 =&0 &\pmod{1} ~. \nonumber
\end{eqnarray}
Here $ V_i \cdot V_j$ is generalized to
\begin{equation}
 V_i \cdot V_j =
  {\vec U}_i \cdot {\vec U}_j -
  \sum_{r} V^r_i V^r_j
  +\sum_{\ell: ~right} T^\ell_i T^\ell_j
  -\sum_{\ell: ~left} T^\ell_i T^\ell_j ~.
\end{equation}
The spectrum generating formula reads:
\begin{equation}\label{sgfz}
 V_i \cdot {\cal N}_{\overline {\alpha V}} +f_i ({\overline {\alpha V}},
 {\bf n})
 +{{\alpha^* t^*}\over{2t_i}}{\vec Q}^2({\vec P}_{\overline {\alpha V}})
 =\sum_j k_{ij} \alpha_j + s_i -V_i \cdot ({\overline {\alpha V}} -
 W({\overline {\alpha V}}))
 \pmod{1} ~.
\end{equation}
Here, ${\vec Q}({\vec P}_{\overline {\alpha V}})
\in {\tilde I}({\overline {\alpha V}})$ is an arbitrary vector such that
${\vec P}_{\overline {\alpha V}}-\alpha^*
{\vec Q}({\vec P}_{\overline {\alpha V}})
\in I({\overline {\alpha V}})$. For example, for $\alpha^*=1$, we can choose
${\vec P}_{\overline {\alpha V}}={\vec Q}({\vec P}_{\overline {\alpha V}})$.
\begin{equation}\label{dotproduct1}
 V_i \cdot {\cal N}_{\overline {\alpha V}} \equiv
  {\vec U}_i \cdot {\vec P}_{\overline {\alpha V}} -
 \sum_{r} V^r_i N^r_{\overline {\alpha V}}
 + \sum_{\ell: ~right} T^\ell_i J^\ell_{\overline {\alpha V}}
 - \sum_{\ell: ~left} T^\ell_i J^\ell_{\overline {\alpha V}} ~.
\end{equation}
where $N^r_{\overline {\alpha V}}$ are the fermion number operators, and 
$J^\ell_{\overline {\alpha V}}$ the boson number operators. Both have integer
eigenvalues.

{}The states that satisfy the spectrum generating formula include
both on- and off-shell states.
The on-shell states must satisfy the additional constraint that
the left- and right-moving energies are equal. In the
${\overline {\alpha V}}$ sector they are given by:
\begin{eqnarray}\label{LRenergy}
 E^L_{\overline {\alpha V}} =&&-1 +\sum_{\ell:~left}
 \{ {1\over 2} {\overline {\alpha T}}^{\,\ell} (1-{\overline
 {\alpha T}}^{\,\ell})+
 \sum_{q=1}^{\infty} [(q+{\overline {\alpha T}}^{\,\ell} -1)n^\ell_q +
(q-{\overline {\alpha T}}^{\,\ell} ){\overline n}^{\,\ell}_q ]\} \nonumber\\
 &&+\sum_{q=1}^{\infty} q (n^0_q +{\overline n}^{\,0}_q )+{1\over 2}
 ({\vec P}^L_{\overline {\alpha V}}+\alpha{\vec U}^L)^2 ~,\\
 E^R_{\overline {\alpha V}} =&&-{1\over 2}+\sum_{\ell:~right}
 \{ {1\over 2} {\overline {\alpha T}}^{\,\ell} (1-{\overline {\alpha T}}
 ^{\,\ell})+
 \sum_{q=1}^{\infty} [(q+{\overline {\alpha T}}^{\,\ell} -1)m^\ell_q +
 (q-{\overline {\alpha T}}^{\,\ell} ){\overline m}^{\,\ell}_q ]\} \nonumber\\
 &&+\sum_{q=1}^{\infty} q (m^0_q +{\overline m}^{\,0}_q )+{1\over 2}
 ({\vec P}^R_{\overline {\alpha V}}+\alpha{\vec U}^R)^2 \nonumber\\
 &&+\sum_{r}
 \{ {1\over 2} ({\overline {\alpha V}}^{\, r})^2+
 \sum_{q=1}^{\infty} [(q+{\overline {\alpha V}}^{\, r} -{1\over 2} )k^r_q +
 (q-{\overline {\alpha V}}^{\, r}-{1\over 2 }){\overline k}^{\, r}_q ]\} ~.
\end{eqnarray}
Here $n^\ell_q$ and ${\overline n}^{\,\ell}_q$ are occupation numbers for the
left-moving bosons $\phi^\ell_L$,
whereas $m^\ell_q$ and ${\overline m}^\ell_q$ are those
for the right-moving bosons
$\phi^\ell_R$. These take non-negative integer values.
$k^r_q$ and ${\overline k}^{\, r}_q$ are the occupation numbers for the
right-moving fermions, and they take only two values: $0$ and $1$.
The occupation numbers are directly related to the boson and fermion number
operators. For example, $N^r_{\overline {\alpha V}} =\sum_{q=1}^{\infty}
(k^r_q -{\overline k}^{\, r}_q)$.

{}To obtain non-abelian orbifold models, with the non-abelian group, 
we consider the abelian orbifolds,
whose abelian groups are subgroups of the non-abelian group. 
We follow the same approach given in section III to obtain the consistency
constraints on the sets of $\{ k_{ij}, V_i \}$. 
Since the derivation and the rules are identical,
we shall not repeat it here. 

\section{Examples}

{}The simpliest example is the ${\bf Z}_2$ orbifold of a single boson. 
This well known example can also be recast in this formalism 
and is given in Appendix A.
Here we give two orbifold models. The first model has the permutation group 
$S_3$ as the non-abelian point group. The second one is simply the
$3$-family $SO(10)_3$ grand unification model given in Ref.\cite{three}, 
recast in the non-abelian orbifold language. Its isometry group is
$T\times {\bf Z}_2$, where T is the tetrahedral group.

%\hfill
\bigskip
\noindent {\em (1) The $S_3$ Model}
\smallskip

{}Let us start with the following 4-dimensional Narain model, with 
$\Gamma^{6,22}=\Gamma^{6,6} \otimes \Gamma^{16}$, where $\Gamma^{16}$ is the 
${\mbox{Spin}}(32)/{\bf Z}_2$ lattice, and
$\Gamma^{6,6}=\left( \Gamma^{1,1} \otimes \Gamma^{2,2} \right)^{2}$ where
both $\Gamma^{1,1}$ and $\Gamma^{2,2}$ are specific even self-dual lattices:
$\Gamma^{1,1}=\{ (p_R \vert\vert p_L) \mid 
~p_R = {1 \over \sqrt{2}} (m-n), ~p_L= {1 \over \sqrt{2}} (m+n) 
~ ; ~ m,n \in \bZ \}$,
{\it i.e.}, an $SU(2)$ lattice and
$\Gamma^{2,2}=\{ (\vec{p}_R \vert\vert \vec{p_L}) \mid
~\vec{p}_{R,L}= {1\over {2R}} \tilde{e}^{i}m_{i} \mp 
R e_{i}n^{i} \}$
where $e_i$ and $\tilde{e}^{i}$ are the $SU(3)$ simple roots and their duals
and $R$ is the compactification radius.
This $N=4$ model (called $N4$) has  
$SU(2) \times SU(2) \times U(1)^{2} \times U(1)^{2} \times SO(32)$
gauge group.

{}Let us first consider the ${\bZ}_3$ orbifold:
\begin{eqnarray} \label{S3}
V_0 &=& (\hf (\hf~0)^{2}_r~(\hf~0)^{2} \vert\vert 0^{2}_r~0^2 \vert 0^{16}_r)
 ~, \nonumber \\
V_1 &=& ( 0 (0~0)^{2}_r ~(-{1\over 3}{1\over 3})({1\over 3}~{2\over 3}) 
\vert\vert (-{\sqrt{2}\over 3})_r({\sqrt{2}\over 3})_r ~{1\over 3}~{2\over 3}
\vert ({1\over 3})^8_r ~({2\over 3})^2_r ~0^6_r ) ~, \\
W_1 &=& ( 0 (0~0)^{2}_r ~(0 ~\phf)^2 \vert\vert 0^2_r ~(\phf)^2 
\vert 0^{16}_r ) ~, \nonumber
\end{eqnarray}
where the subscript $r$ labels real fermions and/or bosons.
Since this is a symmetric ${\bZ}_3$ twist, $V_1 \cdot W_1 =0$,
\begin{equation}
V_i \cdot V_j = \left( \begin{array}{rr}
                 -1 & 0 \\
                 0  & 2 
                         \end{array}
                  \right) ~, \\
\quad \quad
k_{ij}=0 ~.
\end{equation}
Using the spectrum generating formula (\ref{sgfz}), it is straightforward
to work out the massless spectrum of this model.
This orbifold yields a $N=2$ model (called $S3$) which has the
gauge symmetry
$U(1) \times U(1) \times U(10) \times SO(12)$. Here, 
the $U(1)$ factors come from $\Gamma^{1,1}$s.

{}Next we consider a ${\bZ}_2$ model. Let us consider the two bosons $\phi_1$,
$\phi_2$ in $\Gamma^{2,2}$. Under this ${\bZ}_2$ twist,
\begin{eqnarray}
\phi_{2} &\rightarrow& - \phi_2  \nonumber \\
\phi_{1} &\rightarrow&  \phi_1 + \phi_2 
\end{eqnarray}
{\it i.e.}, it is a reflection that leaves $2 \phi_1 + \phi_2$ invariant
(Note that since this reflection acts on both left- and right-moving momenta,
it can be viewed as a rotation in a four dimensional {\em Euclidean} 
lattice spanned by the vectors $(p_R \vert\vert p_L)$. In the example below,
we could alternatively combine two reflections acting on the left-moving 
momenta into a ${\bf Z}_2$ rotation). In 
fact,
$\phi_2$ and ${1\over \sqrt{5}}~(2 \phi_1 + \phi_2)$ form the diagonal basis 
of this ${\bZ}_2$ twist.
Using this basis for both $\Gamma^{2,2}$, the ${\bZ}_2$ orbifold is
given by
\begin{eqnarray}
V_0^{'} &=& (\hf(\hf~0)^{6}_r \vert\vert 0^{6}_r \vert 0^{16}_r ) 
~, \nonumber \\
V_1^{'} &=& (\hf(\hf~\phf)^{2}_r ~[(\hf~\phf)_r(00)_r ]^{2} \vert\vert
(\phf)_r~(\phf~0)^{2}_r \vert (\phf)^{12}_r~0^4_r ) ~, \\
W_1^{'} &=& (0~(0~\phf)^{2}_r~[ (0~\phf)_r~(0~0)_r ]^{2} \vert\vert
(\phf)^{2}~(\phf~0)^{2}_r \vert(\phf)^{12}_r~0^4_r) ~. \nonumber
\end{eqnarray}
Here, $V_1^{'} \cdot W_1^{'} ={3 \over 2}$,
\begin{equation}
V_i^{'} \cdot V_j^{'} =  \left( \begin{array}{rr}
                 -1 & \hf \\
               \hf  & 1 
                         \end{array}
                  \right) ~, \\
\quad \quad
k_{ij}^{'}            = \left( \begin{array}{cc}
                                0 & k_{10}^{'} + \phf \\
                      k_{10}^{'}  & k_{10}^{'}
                                 \end{array}
                          \right) ~.
\end{equation}
Here the two $\Gamma^{1,1}$ as well as $\Gamma^{16}$ are twisted.
This ${\bZ}_2$ orbifold yields a $N=2$ model (called $S2$) with gauge
symmetry $U(1)^{4} \times SO(12) \times SO(20)$.

{}A comment on our choice of $\Gamma^{2,2}$ is necessary here. 
The invariant sublattice $I$ of $\Gamma^{2,2}$ under the ${\bf Z}_3$
twist consists of the origin $(\vec{0} \mid\mid \vec{0})$ only.
On the other hand, the invariant sublattice $I^{'}$ of $\Gamma^{2,2}$
under the ${\bf Z}_2$ twist is non-trivial:
\begin{equation}
I^{'} = \{ {m \over{2R}} ~( \tilde{e}^{1} \mid\mid \tilde{e}^{1} )
+ 3Rn ~( - \tilde{e}^{1} \mid\mid \tilde{e}^{1} )~;~m,n \in {\bf Z} \}
\end{equation}
where $\tilde{e}^{1}$ is the dual of $e_1$. Here $I^{'}$
is even but not self-dual. It is easy to find its dual $\tilde{I}^{'}$,
\begin{equation}
\tilde{I}^{'} = \{ {m \over{4R}} ~( \tilde{e}^{1} \mid\mid \tilde{e}^{1} )
+ {3\over 2} Rn ~(- \tilde{e}^{1}\mid\mid \tilde{e}^{1} )~;~m,n \in {\bf Z} 
\} ~.
\end{equation}
Recall $N_{I^{'}}$, where $N_{I^{'}} \vec{P}^{2} \in 2 {\bf Z}$ for 
$\vec{P} \in \tilde{I}^{'}$. In this case, $N_{I^{'}}=t_1^{'}=2$, which is
consistent. If we turn on appropriate antisymmetric background field and
choose $R=2$ to get an enhanced $SU(3)$ gauge symmetry, we will find
that the corresponding $N_{I^{'}}=4$, which is unacceptable according 
to our rules. In fact, we must have zero antisymmetric background field. 
As a consequence, we have only $U(1)^{2}$ gauge symmetry
from each $\Gamma^{2,2}$.

{}Notice that the ${\bZ}_3$ twist and the ${\bZ}_2$ twist on $\Gamma^{2,2}$
do not commute; in fact, they generate the permutation group $S_3$ as the 
point group. So the final orbifold involves a symmetric $S_3$ and 
an asymmetric ${\bZ}_2$ twists.
Since the ${\bZ}_3$ shifts and the ${\bZ}_2$ twist on part of 
$\Gamma^{16}$ do not commute, they generate the isometry group $D_3$,
which is the same as $S_3$. 
The group $S_3$ has 2 generators $\theta$ and $r$ satisfying $\theta^3=r^2=1$
and $r \theta^{2}=\theta r$. It is divided into 3 conjugacy classes: 
$C_{1}=\{1\}$, $C_{\theta}=\{\theta, \theta^{2}\}$ and 
$C_{r}=\{r, r\theta, r \theta^{2}\}$. The stabilizer groups are given by
$N_{1}=S_3$, $N_{\theta}=\{1,\theta,\theta^{2}\}$ and $N_{r}=\{1,r\}$. 
It is straightforward to check that the non-abelian constraints given in 
Section III are easily satisfied. 
The resulting orbifold has the partition function
\begin{equation}
Z = \phf ~Z_{S3} + Z_{S2} - \phf ~Z_{N4} ~.
\end{equation}
It is not difficult to work out the final massless spectrum of this $N=1$ 
supersymmetric model. The gauge symmetry, coming only from $\Gamma^{16}$, 
is $U(1) \times SO(10)_2 \times SO(10)_1$, where the subscripts indicate
the levels of the current algebra $SO(10)$.

\bigskip
\noindent {\em (2) The $3$-family $SO(10)_3$ model\/}\cite{three}
\smallskip

{}We start with a $N=4$ supersymmetric
Narain model, which was referred to as $N0$,
with the lattice $\Gamma^{6,22}=\Gamma^{2,2} \otimes
\Gamma^{4,4} \otimes \Gamma^{16}$. Here $\Gamma^{2,2} =\{(p_R
\vert\vert p_L ) \}$ is an even self-dual Lorentzian lattice with
$p_R ,p_L \in {\tilde
\Gamma}^2$ ($SU(3)$ weight lattice), $p_L - p_R \in \Gamma^2$
($SU(3)$ root lattice). Similarly, $\Gamma^{4,4} =\{(P_R
\vert\vert P_L ) \}$ is an even self-dual Lorentzian lattice with
$P_R ,P_L \in {\tilde
\Gamma}^4$ ($SO(8)$ weight lattice), $P_L - P_R \in \Gamma^4$
($SO(8)$ root lattice). $\Gamma^{16}$ is the ${\mbox{Spin}}(32)/{\bf
Z}_2$ lattice. This model has $SU(3) \times SO(8) \times SO(32)$ gauge group.

{}Next, consider the model generated by the following vectors:
\begin{eqnarray}
 &&V_0 =(-{1\over 2} (-{1\over 2}~ 0)^3 \vert\vert 0^3 \vert 0^{16}_r ) ~,
 \nonumber \\
 \label{Wil3} 
 &&V_1 =( 0 (0~{1\over 2}e_1 )(0~a_1)(0~b_1) \vert\vert 0^3 \vert ({1\over
 2})^5_r ~0^5_r ~0^5_r ~(-{1\over 2})_r )~,\\
 &&V_2 =( 0 (0~{1\over 2}e_2 )(0~a_2)(0~b_2) \vert\vert 0^3 \vert 0^5_r 
 ~({1\over 2})^5_r ~0^5_r ~(-{1\over 2})_r )~. \nonumber
\end{eqnarray}
Here $e_1$ and $e_2$ are the simple roots of $SU(3)$ ($e_1 \cdot e_1 =
e_2 \cdot e_2 = -2 e_1 \cdot e_2 = 2$), whereas the four-dimensional real
vectors
${\bf s}=(a_1 ,b_1)$ and ${\bf c}=(a_2 ,b_2)$ are respectively
the spinor and conjugate weight vectors of $SO(8)$. The generating vectors
$V_1$ and $V_2$ are order two shifts ($m_1 =m_2 =2$). They play
the role of Wilson lines. The auxiliary vectors
$W({\overline {\alpha V}})$ are therefore null as there is no twisting of the
lattice. 

{}The matrix of the dot products $V_i \cdot V_j$ reads
\begin{eqnarray}
 V_i \cdot V_j &=&\left( \begin{array}{ccc}
                 -1 & 0 & 0\\
                 0  & 0 & 1\\
                 0  & 1 & 0
               \end{array}
        \right) ~,\\
 k_{ij}&=&0 ~.
\end{eqnarray}
{}The resulting model, which was referred to as $N1$,
is a Narain model with $N=4$ space-time supersymmetry and the gauge
group $SU(3)\times SO(8) \times SO(10)^3 \times SO(2)$. The details of 
this construction can be found in Ref.\cite{Kaku,three}.

Now, we consider the asymmetric
orbifold model generated by the following vectors:
\begin{eqnarray}\label{B3}
 V_0 &=&(-{1\over 2} (-{1\over 2}~ 0)^3 \vert\vert 0^3 \vert 0^{5}~0_r^5 ~0_r) 
 ~, \nonumber \\
 V_1^{'} &=&( 0 (-{1\over 3}~{1\over 3})^3 \vert\vert 0~({1\over 3})^2 \vert
 ({1\over 3})^5 ~0_r^5 ~({2\over 3})_r ) ~, \\
 W_1^{'} &=&( 0 (0~{1\over 2})^3 \vert\vert 0~({1\over 2})^2 \vert
 ({1\over 2})^5 ~0_r^5 ~0_r ) ~. \nonumber
\end{eqnarray}
The matrix of the dot products $V_i^{'} \cdot V_j^{'}$ reads
\begin{equation}
 V_i^{'} \cdot V_j^{'}=\left( \begin{array}{rr}
               -1 & \hf \\
               \hf & -{1 \over 3} 
               \end{array}
        \right) ~.
\end{equation}
The structure constants $k_{ij}$ are then given by
\begin{equation}
 k_{ij}=\left( \begin{array}{cc}
               0 & 0 \\
               \phf & {1 \over 3} 
               \end{array}
        \right) ~.
\end{equation}

{}Suppose we start from the $N1$ model, with the vectors (\ref{B3})
acting on it. This is the abelian orbifold construction.
Here the right- and left-moving bosons corresponding to the $SU(3)\times
SO(8)$ subgroup are complexified, whereas the single boson corresponding to
the $SO(2)$ subgroup is real, so are the five bosons $\varphi^I \equiv {1\over
\sqrt{3}}(\phi^I_1 +\phi^I_2 +\phi^I_3)$; the other ten real bosons are
complexified via linear combinations $\Phi^I \equiv {1\over
\sqrt{3}}(\phi^I_1 +\omega\phi^I_2 +\omega^2 \phi^I_3)$ and
$(\Phi^I)^\dagger \equiv {1\over
\sqrt{3}}(\phi^I_1 +\omega^2\phi^I_2 +\omega \phi^I_3)$, where $\omega =\exp
(2\pi i /3)$ (The real bosons $\phi^I_p$, $I=1,...,5$, correspond to the
$p^{\mbox{th}}$ $SO(10)$ subgroup, $p=1,2,3$). Thus, 
the orbifold group element described by the $V_1$ vector 
permutes the real bosons corresponding to the three $SO(10)$s, which in turn
is equivalent to modding out by their outer automorphism.
The resulting model (called $A1$) is simply the $SO(10)$ 
grand unification, with 
gauge symmetry $SU(3)_1 \times SU(3)_3 \times SO(10)_3 \times U(1)$ and
$9$ chiral $SO(10)$ families.
The details can be found in Ref.\cite{Kaku,three}.

{}Notice that the world-sheet basis that is diagonal under
the action of either of the Wilson lines (\ref{Wil3}) 
is different from the basis that is
diagonal under the ${\bf Z}_3$ twist. This implies that the ${\bf Z}_3$ 
twist action does not commute with the Wilson line action.
So if we want to start from the
original $N0$ model to reach the $A1$ model, we must perform a
non-abelian orbifold, {\it i.e.}, an isometry $\overline{P}$ composed of the 
Wilson lines (\ref{Wil3}) and the ${\bf Z}_3$ twist plus shifts (\ref{B3}). 
Since the Wilson lines
correspond to a ${\bf Z}_2 \otimes {\bf Z}_2$ (shift) action, we see that
$\overline{P}=T$, the tetrahedral group. 

{}Let the resulting model coming from the above ${\bf Z}_3$
orbifold ({\it i.e.}, ${V_0,V_1'}$ of Eq(\ref{B3})) on the original $N0$ 
model be referred to as the $B1$ model. Again, it is easy to work out the 
massless spectrum of the 
$B1$ model. It has $N=1$ supersymmetry and gauge symmetry 
$SU(3)_1\times SU(3)_3 \times SU(11)_1\times SO(10)_1\times U(1)$. 
Besides the usual gravity and gauge supermultiplets, this $B1$ model also 
has, in its massless part of the spectrum, three copies of 
$({\bf 1}, {\bf 10}, {\bf 1}, {\bf 1})(0)_L$, 
$({\bf 1}, {\bf 1}, {\bf 1}, {\bf 45})(0)$, 
$({\bf 1}, {\overline {\bf 3}}, {\bf 1}, {\bf 16})(-1)_L$, 
$({\bf 1}, {\overline {\bf 3}}, {\bf 1}, {\bf 10})(+2)_L$, and
$({\bf 1}, {\overline {\bf 3}}, {\bf 1}, {\bf 1})(-4)_L$. We see that
all the massless states are singlets of $SU(3)_1\times SU(11)_1$.
Note that the $SU(3)_3$ chiral anomaly of the chiral field
$({\bf 1}, {\bf 10}, {\bf 1}, {\bf 1})(0)_L$ is
cancelled by that of the other fields, as a
${\bf 10}_L$ of $SU(3)$ has $27$ times the anomaly contribution
of a ${\bf 3}_L$. The model is also $U(1)$ anomaly-free
due to the underlying $E_6$ structure of the $SO(10)\times U(1)$
matter fields as can be seen from the branching
${\bf 27} ={\bf 16}(-1) +{\bf 10} (+2) +{\bf 1}(-4)$
under $E_6 \supset SO(10) \times U(1)$.

{}Now the $A1$ model may be reached as the non-abelian orbifold of the $N0$
model, whose partition function is a linear 
combination of those of the $N0$, the $N1$
and the $B1$ models. It is easy to see that the non-abelian rules given in
Section III are satisfied. The tetrahedral group $T$ has 12 elements
generated by $r$ and $\theta$ satisfying $r^2=\theta^3=(r \theta)^3=1$.
There are 4 conjugacy classes: $C_{1}=\{1\}$, $C_{r}=\{r,r^{'},rr^{'}\}$,
$C_{\theta}=\{\theta, r \theta, r^{'} \theta, rr^{'} \theta \}$ and
$C_{\theta^2}=\{\theta^2, (r \theta)^2, (r^{'} \theta)^2, (rr^{'}\theta)^2\}$
where $r^{'}=\theta r \theta^{-1}$. The stabilizer groups are given by
$N_{1}=T$, $N_r=\{1, r, r^{'}\}$ and $N_{\theta}=\{1, \theta, \theta^{2}\}$.
So the partition function of the $A1$ model is given by
\begin{equation}
Z_{A1}= Z_{B1} +  {1 \over 3}~Z_{N1}  - {1 \over 3}~Z_{N0} ~.
\end{equation}
As mentioned above, the $A1$ model has gauge symmetry
$SU(3)_1 \times SU(3)_3 \times SO(10)_3 \times U(1)$.
Besides the usual gravity and gauge supermultiplets, the 
massless spectrum of the resulting $A1$ model has
three copies of
$({\bf 1}, {\bf 10}, {\bf 1})(0)_L$,
$({\bf 1}, {\bf 1}, {\bf 45})(0)$,
$({\bf 1}, {\overline {\bf 3}}, {\bf 16})(-1)_L$,
$({\bf 1}, {\overline {\bf 3}}, {\bf 10})(+2)_L$, and
$({\bf 1}, {\overline {\bf 3}}, {\bf 1})(-4)_L$.
Comparing the $A1$ and the $B1$ models, it seems that the Wilson lines
(\ref{Wil3}) removes the $SU(11)$ gauge bosons in the $B1$ model 
while raising the $SO(10)$ 
current algebra level from $1$ to $3$. This happens because the 
${\bf Z}_3$ twist on the $SO(32)$ in the $N0$ model is an inner-automorphism.
So this ${\bf Z}_3$ twist is equivalent to some ${\bf Z}_3$ shifts,
which cannot change the current algebra level or cut the rank of the 
gauge group. On the other hand, 
the same ${\bf Z}_3$ twist on the $SO(10)^3$ in the
$N1$ model is an outer-automorphism, which converts $SO(10)^3$ to $SO(10)_3$.

{}In the above example, one may prefer the two-step abelian orbifold 
approach over the non-abelian orbifold approach, since starting 
from the $N1$ model is almost as easy as starting from the $N0$ model.
However, in the case where the point group is non-abelian, such as the 
$S_3$ model discussed earlier, we do not see 
an obvious alternative. 

{}To reach the final $3$-family $SO(10)$ model, one must reduce the 
number of chiral families from $9$ to $3$ by further orbifolding
the $A1$ model by a ${\bf Z}_2$ twist. This ${\bf Z}_2$ orbifold is 
generated by the following vectors:
\begin{eqnarray} 
 V_0 &=&(-{1\over 2} (-{1\over 2}~ 0)^3 \vert\vert 0^3 \vert 0^{5}~0^5_r ~0_r) 
 ~, \nonumber \\
 \label{BC2}
 V_2^{'} &=&( 0 (0~0)(-{1\over 2}~{1\over 2})^2 \vert\vert ({1\over 2}e_1)
 ({1\over 2})^2 \vert 0^5 ~0^5_r ~0_r ) ~, \\
 W_2^{'} &=&( 0 (0~0)(0~{1\over 2})^2 \vert\vert 0
 ({1\over 2})^2 \vert 0^5 ~0^5_r ~0_r ) ~. \nonumber
\end{eqnarray}
The final model is the $3$-family $SO(10)$ grand unified model 
presented in Ref.\cite{three}, with gauge symmetry 
$SU(2)_1 \times SU(2)_3 \times SO(10)_3 \times U(1)^3$. 
Now, this ${\bf Z}_2$ twist commutes with
the ${\bf Z}_3$ twist (\ref{B3}) and the Wilson lines (\ref{Wil3}). So, 
starting from the $N0$ model, the orbifold group is $T \times {\bf Z}_2$. 

{}There are alternative routes to the same final $3$-family $SO(10)$ model. 
We may start with the ${\bf Z}_2$ orbifold (\ref{BC2})
of the $N0$ model, which we refer to as the $N2$ model. 
If we now orbifold this $N2$ model by the tetrahedral group
$T$, the partition function of the final model is given by 
\begin{equation}
Z_{F}= Z_{A3} +  {1 \over 3}~Z_{N3}  - {1 \over 3}~Z_{N2}
\end{equation}
where $N3$ is the ${\bf Z}_2 \times {\bf Z}_2$ orbifold 
({\it i.e.}, $\{V_0, V_1, V_2\}$) of the $N2$ model, and  
$A3$ is the ${\bf Z}_3$ orbifold (\ref{B3})
of the $N2$ model. The final model is again the same three-family $SO(10)$
grand unified model.
 
\acknowledgements

This work was supported in part by the National Science Foundation.

\hfill 

\appendix \section{The ${\bZ}_2$ Orbifold of a Single Boson}
\bigskip

{}The ${\bZ}_2$ orbifold of a single boson compactified at 
arbitrary radius is well understood. In this case, the orbifold involves
a non-abelian isometry group. It is therefore a good example 
to illustrate how the rules for constructing free fermionic string 
models and general abelian orbifolds are generalized to include 
the construction of non-abelian orbifolds. 
In particular, it will help clarify our notation.
This example also illustrates the following point: how the same
model $\Omega$ can be reached via an abelian or a non-abelian orbifold
depending on the model $\Omega^{'}$ that we start with. 

{}In the bosonic formulation, one can start with a torus $\Gamma$
of any desired radius and identify points on $\Gamma$ under 
an abelian point group $P={\bZ}_2$. We can also start with a torus 
$\overline{\Gamma}$ of fixed radius $1$ and obtain the same model by 
the combined action of (1) changing the radius, and (2) identifying points 
on the new torus $\Gamma$ under $P={\bZ}_{2}$. These two operations 
do not commute and generate the non-abelian isometry group $\overline{P}$.

{}To be explicit, let us recall that $\overline{P}$ is a subgroup of 
the space group $S$ 
of rotations and translations. An element of $\overline{P}$ therefore takes 
$X \rightarrow \theta X + v$ and is denoted by $(\theta,v)$. 
The multiplication law for elements of $\overline{P}$ is given by 
$(\theta,v)(\omega,u)=(\theta \omega,v+\theta u)$.

{}As an example, let us construct the ${\bZ}_{2}$ orbifold at radius 
${1 \over n}$ (for simplicity, we take $n$ to be odd). 
Starting from a torus of radius $1$, we can change the radius to ${1\over n}$
by modding out the lattice by a ${\bZ}_n$ shift generated by 
$p=(1,{1\over n})$.
On the other hand, modding out the torus by a twist $q=(-1,0)$
results in a ${\bZ}_2$ orbifold. It is easy to verify 
that $p^n=q^2=1$ and $pq=qp^{n-1}$. Therefore, the group generated by 
$p$ and $q$ is $D_{n}$ which is divided into $\phf (n+3)$ conjugacy classes: 
$C_{p^m}=\{ p^m , p^{-m} \}$ for 
$m=0,\cdots,\phf~(n-1)$, and $C_{q}= \{ q,pq,\cdots p^{n-1}q \}$.
The stabilizer groups are $N_{p^m}=\{ 1,p,\cdots, p^{n-1} \}$ and
$N_{p^{m}q}=\{ 1, p^{m}q \}$ for $m=0,\cdots,n-1$. 
Therefore, the full partition function can be expressed as a sum of the 
three abelian orbifolds, 
\begin{eqnarray}\label{D_n}
Z &=& \phf~Z_p + {1\over n}~\sum_{m=0}^{n-1} Z_q^{m} -\phf~Z_1 \nonumber \\
  &=& \phf~Z_p + Z_q -\phf~Z_1
\end{eqnarray} 
where $Z_p$ is the partition function of a single boson at radius $1\over n$,
each $Z_q^{m}$ is equal to $Z_q$, a ${\bZ}_2$ orbifold partition 
function at radius $1$,
and $Z_1$ is the original
partition function of a single boson at radius $1$, {\em i.e.}, 
that of the starting model $\Omega^{'}$.

{}We now turn to the free fermionic construction. Consistency requires the
presence of $V_0$. In the case of one complex world-sheet fermion,
we always start from the model $\Omega^{'}$, which is equivalent to 
a single boson at radius $1$. Its partition function $Z_{1}$ is 
generated by a single vector
\begin{equation}
V_0 = (\hf \mid \!\hf) ~.
\end{equation}
It is easy to change 
the radius to $1\over n$ by introducing an additional basis vector 
$V_p$. The model $Z_p$ generated by
\begin{eqnarray}
V_0&=&(\hf \mid\!\hf) ~, \nonumber \\
V_p&=&({1\over n} \mid {1\over n}) ~,
\end{eqnarray}
corresponds to a single boson at radius $1 \over n$. We can also construct a
${\bZ}_2$ orbifold by assigning different boundary conditions to the two real
world-sheet fermions. The model $Z_q$ generated by
\begin{eqnarray}
V_0 &=& ((\hf)^{2} \mid \!(\hf)^{2})_r ~, \nonumber \\
V_q &=& (\hf~0 \mid \!\hf~0)_r ~,
\end{eqnarray}
is a ${\bZ}_2$ orbifold at radius $1$. However, the vectors 
$V_p$ and $V_q$ are not compatible because $V_p$ 
must be in the complex fermion basis while $V_q$ is in the real 
fermion basis. So the rules given in Ref.\cite{KLT,KLST} cannot give
the ${\bZ}_2$ orbifold at radius $1 \over n$. To construct this model,
we must use the non-abelian orbifold rules given in Section III.
>From Eq.~(\ref{c1}), we see that $k_{ij}^{q,m}$ are the same for all $m$.
Eq.~(\ref{c2}) implies that $k_{01}^{p}=k_{10}^{p}=k_{11}^{p}=0$
which are automatically satisfied. There is no new constraint from
Eq.~(\ref{c3}) and Eq.~(\ref{h1}). Therefore, $V_p$ and $V_q$
define a consistent model, and the resulting $Z$ is given by Eq.~(\ref{D_n}). 

{}By now, it is almost trivial to write this in the bosonic formulation. 
Let us start with the single boson at radius $1$ ~({\it i.e.} model $Z_1$).
In the real boson notation, the $Z_p$ model is given by the following 
set of vectors acting on the $Z_1$ model,
\begin{eqnarray}
V_0&=&(0 \vert \vert 0)_r ~, \nonumber \\
V_p&=&({1\over n} \vert \vert{1\over n})_r ~, \\
W_p&=&(0 \vert \vert 0)_r ~, \nonumber
\end{eqnarray}
and the $Z_q$ model is given the following
set of vectors acting on the $Z_1$ model,
\begin{eqnarray}
V_0&=&(0 \vert \vert 0)_r ~, \nonumber \\
V_q&=&( \phf \vert \vert \phf)_r ~, \\ 
W_q&=&( \phf \vert \vert \phf)_r ~. \nonumber
\end{eqnarray}
Since this is a symmetric orbifold, the abelian orbifold rules are 
trivial to apply, and we may choose $k_{ij}=0$.
Here, the shift $V_p$ and the twist $V_q$ are the two generators of $D_n$, 
the isometry group ${\overline P}$. 
There is no simultaneously diagonalizable basis
for $V_p$ and $V_q$ because the two operations do not commute. 
In the same fashion, we may apply our non-abelian orbifold rules
in Section III to deduce the consistency constraints on this $D_n$ orbifold.
>From Eq.~(\ref{c1}), we again infer that $k_{ij}^{q,m}$ are the same for 
all $m$. Eq.~(\ref{c2}) implies that $k_{01}^{p}=k_{10}^{p}=k_{11}^{p}=0$
which are also automatically satisfied. There is no new constraint from
Eq.~(\ref{c3}) and Eq.~(\ref{h1}). Therefore, $V_p$ and $V_q$
define a consistent model provided that we interpret the ${\bZ}_n$
shift ($V_p$) and the ${\bZ}_2$ twist ($V_q$) as generators of the
non-abelian group $D_n$. The resulting $Z$ is again given by Eq.~(\ref{D_n}).

%============================================================================
%\newpage
\widetext

\end{document}